\documentclass[12pt,a4paper]{article}

\usepackage[T1]{fontenc}
\usepackage[hmarginratio=1:1,top=32mm,columnsep=20pt]{geometry}
\usepackage{mathdots}
\usepackage{upgreek}
\usepackage{color}
\definecolor{dark-gray}{gray}{0.20}
\definecolor{gray}{gray}{0.30}
\definecolor{light-gray}{gray}{0.80}
\definecolor{dark-red}{rgb}{0.7,0,0}
\definecolor{dark-green}{rgb}{0.1,0.4,0}
\definecolor{dark-blue}{rgb}{0.3,0.3,0.7}
\definecolor{light-blue}{rgb}{0.8,0.8,1}

\usepackage{cite}
\usepackage{hyperref}
\hypersetup{
	colorlinks=true,
	linkcolor=dark-blue,
	citecolor=dark-red,
	urlcolor=dark-green,
	linktoc=page
}

  \usepackage{a4wide}
  \usepackage{latexsym}
  \usepackage{epsf}
  \usepackage{bbm}
  \usepackage{graphicx}
  \usepackage{mathrsfs}
\usepackage{amsmath,amssymb,amsthm}
  \usepackage{verbatim}

\newcommand{\bbm}{\left(\begin{matrix}}
\newcommand{\ebm}{\end{matrix}\right)}
\newcommand{\bea}{\begin{eqnarray}}
\newcommand{\eea}{\end{eqnarray}}
\newcommand{\be}{\begin{equation}}
\newcommand{\ee}{\end{equation}}

\begin{document}
\numberwithin{equation}{section}

\begin{center}

{\LARGE {\bf U-Folds From Geodesics in Moduli Space }}  \\

\vspace{1.5 cm} {\large  D. Astesiano$^a$, D. Ruggeri $^b$ and M. Trigiante$^b$  }\footnote{{ \upshape\ttfamily dastesiano@hi.is, daniele.rug@gmail.com, mario.trigiante@polito.it.
 } }\\
\vspace{1 cm}  ${}^a$ Science Institute, University of Iceland, Dunhaga 3, 107 Reykjav{\'i}k, Iceland,
\\ \vspace{.15 cm}
\vspace{.15 cm} { ${}^b$Department of Applied Science and Technology, Politecnico di Torino, \\ C.so Duca degli Abruzzi, 24, I-10129 Torino, Italy\\ \vspace{.15 cm}}

\vspace{2cm}

{\bf Abstract}

\end{center}
We exploit the presence of moduli fields in the ${\rm AdS}_3\times { S}^3\times CY_2$, where $CY_2=T^4$ or $K3$, solution to Type IIB superstring theory, to construct a U-fold solution with geometry ${\rm AdS}_2\times S^1\times {\rm S}^3\times CY_2$.
This is achieved by giving a non-trivial dependence of the moduli fields in ${\rm SO}(4,n)/{\rm SO}(4)\times {\rm SO}(n)$ ($n=4$ for $CY_2=T^4$ and $n=20$ for $CY_2=K3$ ), on the coordinate $\eta$ of a  compact direction $S^1$ along the boundary of ${\rm AdS}_3$, so that these scalars, as functions of $\eta$, describe a geodesic on the corresponding moduli space. The back-reaction of these evolving scalars on spacetime amounts to a splitting of ${\rm AdS}_3$ into ${\rm AdS}_2\times S^1$ with a non-trivial monodromy along $S^1$ defined by the geodesic. Choosing the monodromy matrix in ${\rm SO}(4,n;\,\mathbb{Z})$, this supergravity solution is conjectured to be a consistent superstring background. We generalize this construction starting from an ungauged theory in $D=2d$, $d$ odd, describing scalar fields non-minimally coupled to $(d-1)$-forms and featuring solutions with topology  ${\rm AdS}_d\times S^d$, and moduli scalar fields. We show, in this general setting, that giving the moduli fields a
geodesic dependence  on the $\eta $ coordinate of an $S^1$ at the boundary of
${\rm AdS}_d$ is sufficient to split this space into ${\rm AdS}_{d-1}\times S^1$, with a monodromy along $S^1$ defined by the starting and ending points of the geodesic. 
This mechanism seems to be at work in the known J-fold solutions in $D=10$ Type IIB theory and hints towards the existence of similar solutions in the Type IIB theory compactified on $CY_2$. We argue that the holographic dual theory on these backgrounds is a 1+0 CFT on an interface in the 1+1 theory at the boundary of the original ${\rm AdS}_3$.

\setcounter{tocdepth}{2}
\newpage
\textcolor{white}{x}\\
\tableofcontents
\newpage
\section{Introduction}
Solutions of superstring theory of $D=11$ supergravity with geometry ${\rm AdS}_d \times M_{{\rm int.}}$ have been the focus of intense study because of their relevance to the ${\rm AdS/CFT}$ correspondence. Type IIB superstring theory, in particular, features a number of backgrounds of the form ${\rm AdS}_d \times S^d\times M_{10-2d}$ which are characterized by moduli fields, i.e. scalar fields which can be assigned an arbitrary constant value throughout spacetime without affecting its geometry. These fields are dual to exactly marginal operators in the dual CFT. The simplest example is ${\rm AdS}_5 \times S^5$ on which the holographic duality was originally conjectured, and whose moduli fields are the Type IIB axion and dilaton fields which correspond to the complexified coupling constant in the dual $\mathcal{N}=4$ SYM theory. Other examples of such backgrounds, or variants thereof, are\footnote{For a review of the backgrounds ${\rm AdS}_3 \times S^3\times CY_2$, with $CY_2=T^4$ or $K3$, and of their holographic dual descriptions, see \cite{Aspinwall:1996mn,David:2002wn,Avery:2010qw}. In the present work, we shall not concentrate on the dual CFT side of these backgrounds.}
\begin{itemize}
    \item[i)] ${\rm AdS}_3 \times S^3\times T^4$ describing the near-horizon geometry of a D1-D5 or of an F1-NS5 system, in which the 5-branes are wrapped around the 4-torus $T^4$;
    \item[ii)] ${\rm AdS}_3 \times S^3\times K3$ describing the near-horizon geometry of a D1-D5 or of an F1-NS5 system, in which the 5-branes are wrapped around the K3 manifold;  
    \item[iii)] ${\rm AdS}_5 \times S^5/\mathbb{Z}_k$ describing the near-horizon geometry of a stack of $D3$ branes at the apex of an orbifold $\mathbb{C}^3/\mathbb{Z}_k$ with $k>1$ \cite{Kachru:1998ys,Katmadas:2018ksp}.
\end{itemize}
The backgrounds  $i)$ can be described within Type IIB superstring theory compactified on $T^4$ to $D=6$, which is described by a  maximal six-dimensional supergravity featuring the global symmetry group ${\rm SO}(5,5)$ at the classical level. The classical moduli space of the ${\rm AdS}_3 \times S^3$ solution of the six-dimensional maximal supergravity is ${\rm SO}(4,5)/{\rm SO}(4)\times {\rm SO}(5)$, ${\rm SO}(4,5)$ being the stabilizer in ${\rm SO}(5,5)$ of the brane charges.\\
The background $ii)$ is a solution to Type IIB superstring theory compactified on K3, which is described, at the classical level, by a half-maximal six-dimensional supergravity and the classical moduli space is ${\rm SO}(4,21)/{\rm SO}(4)\times {\rm SO}(21)$.\\
Finally, in the last case $iii)$, the classical moduli space is  ${\rm SU}(1,k)/{\rm U}(k)$ and acts on the $k$ complexified coupling constants of the dual necklace quiver theory
\cite{Corrado:2002wx,Louis:2015dca},
.\\
All these moduli occur in the $d$-dimensional gauged supergravity featuring the ${\rm AdS}_d$ factor ($d=3$ and 5 in the above examples) as vacuum solution, as flat directions of the scalar potential.\\
Assuming a dependence of these scalar fields on the spacetime coordinates will in general affect the geometry of the background in some non-trivial way. In \cite{Hertog:2017owm,Ruggeri:2017grz,Katmadas:2018ksp,Astesiano:2022qba} the authors have investigated spacetime-dependent configurations of the moduli fields in the Wick-rotated moduli space within the Euclidean version of the theory. The resulting solutions were described by geodesics in the (pseudo-Riemannian) moduli space, which were classified according to their being lightlike, spacelike, and timelike, and an interpretation of the corresponding backgrounds, in the dual CFT, was worked out.\\

Here we shall work in a suitable, ungauged Lorentzian theory in $D=2d$, $d$ odd, exhibiting a ${\rm AdS}_d \times S^d$ background with moduli fields. We show how to construct a ${\rm AdS}_{d-1}\times S^1 \times S^d$ geometry from the ${\rm AdS}_d \times S^d$ one by giving a suitable spatial dependence to the scalar moduli.

The construction proceeds along the following steps. We first compactify one direction in the boundary of ${\rm AdS}_d$. Let us denote by $\eta$ the coordinate of the corresponding $S^1$.
We then assume a suitable subset $\varphi^a$ of the moduli fields to depend on $\eta$ only and to describe, in their evolution, a geodesic in the moduli space as $\eta$ varies in its defining interval of values $\eta\in [0,T]$, $T$ being the length of $S^1$. As a consequence of the back-reaction of the evolving scalar fields on spacetime, the ${\rm AdS}_d \times S^d$ background is {transformed} into an ${\rm AdS}_{d-1}\times S^1 \times S^d$ geometry, with a non-trivial monodromy along $S^1$. The two solutions cannot be continuously deformed into one another. If we denote by $\mathcal{M}_0=G_0/H_0$ the moduli space spanned by $\varphi^a$, and by $O$ and $P$ the starting and ending points of the geodesic, parameterized by $\varphi^a(0)$ and $\varphi^a(T)$, respectively, the $G_0$-element $g$ connecting $P$ to $O$, defines the monodromy: $P=g\,O$. If we assume that $G_0(\mathbb{Z})$ is a symmetry of the underlying superstring theory reduced to ${\rm AdS}_d \times S^d$, then by choosing $g\in G_0(\mathbb{Z})$, the backgrounds defined by the initial and final points of the geodesic would be identified from the string perspective and the corresponding supergravity solution could be conjectured to be a consistent superstring background. The component of the metric in the $S^1$ direction is proportional to the squared "velocity" along the geodesic. Therefore the ${\rm AdS}_{d-1}\times S^1 \times S^d$ is not continuously connected to ${\rm AdS}_{d} \times S^d$ as this velocity is set to zero and the moduli fields to constant values.

A prototype of this background is the non-supersymmetric J-fold \cite{Guarino:2019oct}, with geometry ${\rm AdS}_{4}\times S^1 \times S^5$, characterized by an axio-dilaton field evolving along a geodesic in their moduli space ${\rm SL}(2,\mathbb{R})/{\rm SO}(2)$. In the corresponding J-fold description, the initial and final points of the geodesic are connected by a monodromy $g=J_n\in {\rm SL}(2,\mathbb{Z})_{{\rm IIB}}$, $J_n$ being a hyperbolic element of the Type IIB global symmetry group. 
However, the mechanism analyzed here, of the construction of ${\rm AdS}_{d-1}\times S^1 \times S^d$ U-folds by giving the moduli fields a geodesic dependence on the coordinate of $S^1$, seems to be at work in the  J-fold solutions studied in \cite{Inverso:2016eet,Assel:2018vtq,Guarino:2019oct,Guarino:2020gfe,Giambrone:2021zvp,Guarino:2021kyp,Arav:2021gra,Bobev:2021yya,Cesaro:2021tna,Guarino:2021hrc,Bobev:2021rtg,Guarino:2022tlw} with geometry ${\rm AdS}_{4}\times S^1 \times \tilde{S}^5$, $\tilde{S}^5$ being a deformed 5-sphere. We shall further comment on this in the final Discussion section.\par
In this paper, we shall focus on solutions of the form ${\rm AdS}_{2}\times S^1 \times S^3$ within Type IIB theory compactified on a $CY_2$-manifold, which can either be $T^4$ or $K3$. Such backgrounds feature a richer class of monodromies. In the former case, $CY_2=T^4$, the solutions have an $O(4,4;\mathbb{Z})$-monodromy, in the latter, $CY_2=K3$, the monodromy is chosen in $O(4,20;\mathbb{Z})$. These discrete groups are contained inside the so-called U-duality group which is conjectured to encode all string dualities and to be an exact symmetry of the underlying, though yet unknown, unifying quantum theory of gravity \cite{Hull:1994ys}. We shall work extensively on the $CY_2=T^4$ case and construct explicit U-fold solutions within the effective $\mathcal{N}=(2,2)$ six-dimensional supergravity.

\medskip
The paper is organized as follows. In section \ref{GenConst} we describe the general construction of such ${\rm AdS}_{d-1}\times S^1\times S^d$ backgrounds.
We start from a ${\rm AdS}_d\times S^d$ solution of an ungauged model in $D=2d$, $d$ odd, which describes scalar fields, spanning the symmetric target space of sigma model, and non-minimally coupled to a set of $(d-1)$-forms. We discuss, in subsection \ref{bc}, the issue of the boundary conditions along $S^1$ in relation to the geodesic motion of the moduli fields. In subsection \ref{chis} we also introduce, on such backgrounds, the so-called $\chi$-deformations originally studied, in the context $D=10$ J-fold solutions, in \cite{Giambrone:2021zvp,Guarino:2021kyp,Guarino:2021hrc,Guarino:2022tlw}. To give concrete examples of these solutions, we focus on the Type IIB theory compactified on a 4-torus. We start reviewing, in section \ref{bosonicsector}, the bosonic sector of the resulting ungauged maximal 6-dimensional supergravity. This is the model where we construct explicit U-fold solutions. Then, in section \ref{ReviewAdS3S3} we review the known AdS$_3\times S^3$ solutions and their string interpretation. Starting from these backgrounds, in section \ref{U-Foldsection} we build the U-Fold for a certain choice of the monodromy matrix.

In Appendix A we review the geometric characterization of the Type IIB string origin of the scalar and tensor fields in the maximal six-dimensional supergravity.  In Appendix B we review the general construction of a black-string solution with ${\rm SO}(4)$-symmetry and the attractor mechanism at work for the extremal ones. In Appendix C we review the construction of the D1-D5 solution within the maximal six-dimensional theory, also in the presence of certain, radially evolving,  moduli fields. In Appendix D we give the explicit form of the effective potential in the presence of D1-D5 charges, as a function of the string 0-modes, and discuss its extremization.

\section{The General Construction of the ${\rm AdS}_{d-1}\times S^1 \times S^d$ U-Fold} \label{GenConst}
In this section, we show how to build a new class of solutions with topology ${\rm AdS}_{d-1}\times S^1 \times S^d$ (with dimension $D=2d$) by giving the moduli a suitable non-trivial profile. For the sake of concreteness, we take $d$ odd since we have in mind the type IIB backgrounds with $d=3$ and $5$.\\
We start considering a model, in $D$-dimensions, describing $n$ non-chiral $p$-forms $C^I_{(p)}$ ($I=1,\dots n$), with $p=d-1$, coupled to gravity and to a number of scalar fields $\phi^s$, $s=1,\dots n_s$. The latter are described by a sigma-model with homogeneous symmetric, Riemannian target space:
$${M}_{{\rm scal.}}=\frac{G}{H}\,.$$
We assume the isometry group $G$  of the scalar manifold to admit a pseudo-orthogonal representation $\mathscr{R}_p$ in terms of ${\rm O}(n,n)$-transformations.
Our analysis will be extended to models in which the number $n$ of self-dual and $m$ anti-self-dual $d$-form field strengths of the $p$-form fields are different. In this case, we require $G$ to admit a representation in terms of ${\rm O}(n,m)$-transformations.\par In an extended $D$-dimensional supergravity model, this geometric feature is built-in. As is the case in ungauged $D$-dimensional supergravity, the scalar fields are non-minimally coupled to the tensor ones, that is they couple to their field strengths $H^I_{(d)}\equiv dC_{(p)}^I$ of the latter in their kinetic terms.
The general form of the bosonic action we are considering is
\footnote{Throughout the paper we adopt the mostly plus notation for the metric. Moreover we define \begin{align}\sigma_{(p)}\cdot \omega_{(d)}&\equiv \sigma_{\hat\mu_1\dots \hat\mu_d} \,\omega^{\hat\mu_1\dots \hat\mu_d}/d!\,,\nonumber\\({}^*\omega)_{\hat\mu_1\dots \hat\mu_{D-p}}&\equiv \frac{e_{D}}{p!}\,\epsilon_{\hat\mu_1\dots \hat\mu_{D-p} \hat\nu_1\dots \hat\nu_p}\,\omega^{\hat\nu_1\dots \hat\nu_p}\,,\end{align} 
where $\epsilon_{0,\dots D-1}=1$ and $\star^2\omega_{(p)}=(-1)^{(D-p)p+1}\, \omega_{(p)}$.}
\begin{align}
\mathcal{L}_{2d}= e_{2d} \left[R-\frac{1}{2}\,\partial_{\hat{\mu}} \phi^r \partial^{\hat{\mu}} \phi^s  \mathcal{G}_{rs}(\phi)- \frac{1}{2}\,\mathcal{I}(\phi)_{IJ} H^{I}\cdot H^{J}- \frac{1}{2}\,\mathcal{R}(\phi)_{IJ} {H}^{I} \cdot {}^* H^{J}\right],\label{lagrangian2d}
\end{align}
where $\mathcal{G}_{rs}(\phi)>0$ is the target space metric of the scalar sigma model, and $e_D=e_{2d}\equiv \sqrt{|{\rm det}(g_{\hat{\mu}\hat{\nu}})|}$. The matrices $\mathcal{I}_{IJ}(\phi)>0$ and $\mathcal{R}_{IJ}(\phi)=- \mathcal{R}_{JI}(\phi)$ are functions of the scalar fields and describe the non-minimal coupling between them and the tensor field strengths $H^I$.
 We define the dual field strengths as
\begin{gather}
    G_{I\,\hat{\mu}_1\dots \hat{\mu}_d}=   \epsilon_{\hat{\mu}_1\dots \hat{\mu}_d\hat{\nu}_1..\hat{\nu}_d} \frac{\partial \mathcal{L}_{2d}}{\partial H^{I}_{\,\hat{\nu}_1\dots \hat{\nu}_d}}, 
    \end{gather} 
    which, in matrix form and suppressing the spacetime indices, reads:
    \begin{gather}
    G=-\mathcal{I}\,{}^\star H-   \mathcal{R} \, H\,. 
    \end{gather}
In what follows we treat the field strength and the dual field strength on equal footing and define this column vector \begin{gather}
    \mathbb{H}=(\mathbb{H}^M)= \begin{pmatrix} H^{I} \\ G_{I} \\ \end{pmatrix}.
\end{gather}
From this the following \emph{twisted self-duality condition} follows
 \begin{gather}
   {}^\star \mathbb{H}= - \Omega  \,\mathcal{M}\,\mathbb{H}\,,  \label{SDC1}
\end{gather}
where 
\begin{align}
    \Omega   = \begin{pmatrix}
    0 & \mathbb{I} \\
    \mathbb{I}&0
    \end{pmatrix}\,\,,\,\,\,\mathcal{M}(\phi)\equiv \left(\begin{matrix}
        \mathcal{I}-\mathcal{R}\,\mathcal{I}^{-1}\,\mathcal{R}& -\mathcal{R}\mathcal{I}^{-1}\cr \mathcal{I}^{-1}\mathcal{R} & \mathcal{I}^{-1}
    \end{matrix}\right)\,,\label{matrixM}
\end{align}
 $\Omega $ is the $O(n,n)$-invariant matrix, $\Omega^2=  \mathbb{I}$, and $\mathcal{M}$ is a scalar-dependent pseudo-orthogonal, positive-definite symmetric matrix: $\mathcal{M}\,\Omega\,\mathcal{M}=\Omega$.
The Maxwell equations and the Bianchi equations read,
\begin{align}
    d\mathbb{H}=0 \,. \label{EOMHG} 
\end{align}
The Einstein and scalar field equations of motion for this model are
\begin{align}
      R_{\mu\nu}- \frac{1}{2}g_{\mu\nu}R\,&=\,T^{(s)}_{\mu\nu}+T^{(H)}_{\mu\nu}\,,\nonumber\\
       D_{\hat{\mu}}(\partial^{\hat{\mu}} \phi^s)= \nabla_{\hat{\mu}} (\partial^{\hat{\mu}} \phi^s)+\tilde{\Gamma}^s_{uv}\partial_{\hat{\mu}} \phi^u \partial^{\hat{\mu}} \phi^v&=\frac{1}{4}\,\mathcal{G}^{st}\,\mathbb{H}^T\,\partial_t\mathcal{M}\,\, \mathbb{H}\, \label{ScalarEqGen},
\end{align}
where the energy-momentum tensors are defined as:
\begin{align}
   T^{(s)}_{\hat{\mu}\hat{\nu}}&\equiv  \frac{1}{2}\mathcal{G}_{rs} \left(\partial_{\hat{\mu}} \phi^r \partial_{\hat{\nu}}\phi ^s- \frac{1}{2}g_{\hat{\mu}\hat{\nu}} \partial_{\hat{\rho}} \phi^r \partial^{\hat{\rho}} \phi^s \right)\,,\nonumber\\
   T^{(H)}_{\hat{\mu}\hat{\nu}}&\equiv \frac{1}{4\,p!}\,\mathbb{H}^T_{\hat\mu\hat\mu_1\dots \hat\mu_{p}}\,\mathcal{M}(\phi)\,\mathbb{H}_{\hat\nu}{}^{\hat\mu_1\dots \hat\mu_{p}}\,
\end{align}
and $\tilde{\Gamma}^s_{uv}(\phi)$ denotes the  Levi-Civita connection on the scalar manifold. The scalar kinetic term can also be written in the following form
\begin{gather}
    \frac{{\tt k}}{8}\text{Tr}\left(\mathcal{M}^{-1}\partial \mathcal{M}\mathcal{M}^{-1}\partial \mathcal{M}\right)=\frac{1}{2}\mathcal{G}_{st}\partial_{\hat{\mu}} \phi^s \partial^{\hat{\mu}} \phi^t\,,\label{MdMg}
\end{gather}
where the constant ${\tt k}$ depends of the matrix representation of $\mathcal{M}$.\par
It is straightforward to prove, using the above expression for $T^{(H)}$, that the trace of this tensor vanishes 
  \begin{gather}
    T^{(H)}{}_{\hat{\mu}}{}^{ \hat{\mu}}= \frac{1}{8}\mathbb{H}^T_{\hat{\mu}} \mathcal{M} \mathbb{H}^{\hat{\mu}}= \frac{1}{8} \mathbb{H}^T \mathcal{M} \wedge \star \mathbb{H}=- \frac{1}{8}\mathbb{H}^T \wedge \mathcal{M} \Omega   \mathcal{M}\mathbb{H}=- \frac{1}{8}\mathbb{H}^T \wedge \Omega \mathbb{H}=0\,. \end{gather}
The global symmetry of the above equations in supergravity is the invariance under the two-fold action of the group $G$ on the scalar fields as the isometry group of the scalar manifold, and on the field strengths $\mathbb{H}$ through a pseudo-orthogonal representation $\mathscr{R}_p$ (we suppress all pseudo-orthogonal indices):\footnote{For the sake of simplicity, we use the short-hand notation $A^{-T}\equiv (A^{-1})^T$.}
\begin{align}
    g\in G&\,\,\stackrel{\mathscr{R}_p}{\longrightarrow} \,\,\mathscr{R}_p[g]\in {\rm O}(n,n)\,,\nonumber\\
    \mathbb{H}&\rightarrow\,\, \mathbb{H}'=\mathscr{R}_p[g]\,\mathbb{H}\,,\nonumber\\
    \mathcal{M}(\phi)&\rightarrow\,\,  \mathcal{M}(\phi')=\mathscr{R}_p[g]^{-T}\,\mathcal{M}(\phi)\,\mathscr{R}_p[g]^{-1}\,.\label{dualaction}
\end{align}
In extended supergravities the matrix $\mathcal{M}(\phi)$ obeys the above transformation property since it is expressed in terms of the coset representative $\mathbb{L}(\phi^s)$ of ${M}_{{\rm scal.}}$ in the representation $\mathscr{R}_p$ as follows:\footnote{Here we describe the scalar manifold as a left-coset space. }
\begin{equation}
    \mathcal{M}(\phi)=\mathscr{R}_p[\mathbb{L}(\phi)]^{-T}\,\mathscr{R}_p[\mathbb{L}(\phi)]^{-1}\,.\label{ML}
\end{equation}
The transformation property \eqref{dualaction} of $\mathcal{M}(\phi)$ then follows from the action of an isometry element $g\in G$ on $\phi^s$:
\begin{equation}
g\,\mathbb{L}(\phi)=\mathbb{L}(\phi')\,h(g,\phi)\,\,,\,\,\,h(g,\phi)\in H\,
\end{equation}
and the fact that $\mathscr{R}_p[h(g,\phi)]\in {\rm O}(n)\times {\rm O}(n)$.
We now consider solitonic solutions
described only by the scalar and the tensor fields, with spacetime geometry ${\rm M}_d\times_w S^d$, where we will choose ${\rm M}_d={\rm AdS}_{d-1}\times S^1$. The metric will read
\begin{equation}
ds^2=g_{\mu\nu}\,dx^\mu\,dx^\nu+g_{ij}\,d\upxi^i\,d\upxi^j=v_1^{2}\,ds^2_{{\rm AdS}_{d-1}}+v_3^{2}\,d\eta^2+v_2^{2}\,ds^2_{S^d}\,,
\end{equation}
where $\mu=0,\dots, d-1$ and $i=1,\dots, d$, $ds^2_{{\rm AdS}_{d-1}}$ is the metric of an AdS$_{d-1}$ space of radius 1 and $,ds^2_{S^d}$ is the metric of a $S^d$-sphere of radius 1. The coefficients $v_1,\,v_2,\,v_3$ are constant and we have split the $M_d$ coordinates $(x^\mu)=(x^\alpha,\eta)$, where $\alpha=0\,\dots, d-2$ and $x^{d-1}=\eta$. We shall also define:
$$\tilde{e}_{d}=\sqrt{|{\rm det}(g_{\mu\nu})|}\,\,,\,\,\,{e}_{d}=\sqrt{{\rm det}(g_{ij})}\,.$$ The radius $v_2$ of the $d$-sphere will also be denoted by $L$.\par
The general ansatz for the tensor field strengths reads:
\begin{align}
    \mathbb{H}=- \Omega\,\mathcal{M}\, \Gamma \, \boldsymbol{\epsilon}_{\text{M}_d}  + \Gamma \, \boldsymbol{\epsilon}_{\text{S}^d}, \label{AnsatzH}
\end{align}
where $\boldsymbol{\epsilon}_{{\rm M}}$, $\boldsymbol{\epsilon}_{\text{S}^d}$ denote the volume densities of the manifold ${\rm M}_d$ and $\text{S}^d$:
$$\boldsymbol{\epsilon}_{{\rm M}}\equiv  \frac{\tilde{e}_{d}}{d!\,L^d}\,\epsilon_{\mu_1\dots \mu_d}\,dx^{\mu_1}\wedge \dots\,dx^{\mu_{d}}\,,\,\,\,\boldsymbol{\epsilon}_{\text{S}^d}\equiv \frac{{e}_{d} }{d!\,L^d}\,\epsilon_{i_1\dots i_d}\,d\upxi^{i_1}\wedge \dots\,d\upxi^{i_{d}}\,, $$
where we recall that $L$ is the radius $v_2$ of the $d$-sphere.
We have also denoted by  $\Gamma=(\Gamma^M)$ the pseudo-orthogonal vector of charges:
\begin{equation}
    \Gamma^M\equiv \frac{1}{{\rm S}_{\text{S}^d}}\,\int_{\text{S}^d} \mathbb{H}^M\,,
\end{equation}
where $\text{S}_{\text{S}^d}$ is the surface area of a $d$-sphere of radius 1.
This ansatz is covariant under (\ref{dualaction}) provided the charge vector is transformed accordingly:
\begin{equation}
     g\in G\,\,,\,\,\,\,\Gamma \rightarrow\,\, \Gamma'=\mathscr{R}_p[g]\,\Gamma\,.
\end{equation}
Quantum effects would restrict $\Gamma^M$ to belong to an even, unimodular charge lattice $\boldsymbol{\Gamma}^{n,n}$ and this would, in turn, break the global symmetry group $G$ to a discrete subgroup $G(\mathbb{Z})\sim\mathscr{R}_p[G]\cap {\rm O}(n,n;\mathbb{Z})$ preserving the lattice $\boldsymbol{\Gamma}^{n,n}$. In the effective supergravity description of superstring/M-theories, this group $G(\mathbb{Z})$ was conjectured to encode the known string-dualities \cite{Hull:1994ys}. Field configurations connected by  $G(\mathbb{Z})$ should be identified from the string theory point of view. The values of the $D$-dimensional scalar fields defining inequivalent string backgrounds should then span the manifold:
$$G(\mathbb{Z})\backslash G/H\,.$$
In chiral models, in which the number $n$ of self-dual and the number $m$ of anti-self-dual $d$-form field strengths in $D$-dimensions are different, as is the case of Type IIB superstring theory compactified in $K3$ to $D=6$, 
the duality action of the global symmetry group $G$ on the $d$-form field strengths is implemented by transformations in  ${\rm O}(n,m)$: $\mathscr{R}_p[G]\subset {\rm O}(n,m)$.
Quantum corrections break  ${\rm O}(n,m)$ to ${\rm O}(n,m;\,\mathbb{Z})$ leaving the even, unimodular charge lattice $\boldsymbol{\Gamma}^{n,m}$ invariant. The global symmetry group is consequently broken to $G(\mathbb{Z})$. Our construction also applies to this more general setting.\par
In the remainder of this paragraph, we shall focus on the classical description and thus consider the global symmetry group $G$ to be continuous.\par
As for the space-time dependence of the scalar fields, we need first to define the solution's moduli. We define the \emph{little group}, or \emph{stabilizer}, of the charge vector $\Gamma$ the maximal subgroup $G_\ell$ of $G$ such that:
\begin{equation}
    \forall g\in G_\ell\,\,:\,\,\,\,\mathscr{R}_p[g]\,\Gamma=\Gamma\,.\label{littleG}
\end{equation}
If $H_\ell$ denotes the maximal compact subgroup of $G_\ell$, the moduli fields $\varphi^u$, $u=1,\dots, n_\ell$, are defined as the subset of scalar fields $\phi^s$ spanning the submanifold:
\begin{equation}
   {M}_\ell\equiv \frac{G_\ell}{H_\ell}\,,
\end{equation}
of ${M}_{{\rm scal.}}$ at the origin.
We split the scalar fields accordingly:
$$\{\phi^s\}=(\varphi^u,\,\phi^k)\,\,,\,\,\,u=1,\dots, n_\ell\,\,,\,\,\,k=1,\dots, n_s-n_\ell\,,$$
by defining the coset representative of ${M}_{{\rm scal.}}$ as \cite{Andrianopoli:2010bj}:
\begin{equation}
\mathbb{L}(\phi^s)=\mathbb{L}_\ell(\varphi^u)\,\hat{\mathbb{L}}(\phi^k)\,\,,\,\,\,\mathbb{L}_\ell(\varphi^u)\in \frac{G_\ell}{H_\ell}\,,\label{coset01}
\end{equation}
where the scalars $\hat{\phi}\equiv (\phi^k)$ are defined as follows. 
Consider the Cartan decomposition of the Lie algebra $\mathbb{G}$ of $G$ into the Lie algebra $\mathbb{H}$ generating $H$ and the coset space $\mathbb{K}$ isomorphic to the tangent space to ${M}_{{\rm scal.}}$ at the origin:
\begin{equation}
    \mathbb{G}=\mathbb{H}\oplus \mathbb{K}\,.
\end{equation}
The latter space $\mathbb{K}$ supports a representation of $H$ under its adjoint action. This representation is completely reducible under the adjoint action of $H_\ell\subset H$ and $\mathbb{K}$ splits accordingly in $H_\ell$-invariant subspaces
\begin{equation}\mathbb{K}=\mathbb{K}_\ell\oplus \hat{\mathbb{K}}\,,\label{KKK}\end{equation}
where $\mathbb{K}_\ell$ is the coset space of $G_\ell/H_\ell$, while $\hat{\mathbb{K}}$ supports a representation of $H_\ell$.
We define 
\begin{equation}
\hat{\mathbb{L}}(\phi^k)\in e^{\hat{\mathbb{K}}}\,.
\end{equation}
We note that the scalar fields $\varphi^u$ span a symmetric submanifold 
$G_\ell/H_\ell$ of ${M}_{{\rm scal.}}$ only at $\phi^k=0$. For fixed $\hat{\phi}=(\phi^k)\neq (0)$, $\varphi^u$ span a space which is not a symmetric submanifold $G_\ell/H_\ell$ of ${M}_{{\rm scal.}}$ since this space does not support a transitive action of $G_\ell$. Indeed, using the short-hand notation $\phi\equiv (\phi^s)$ and $\varphi_\ell\equiv (\varphi^u)$, 
the action of an element $g_\ell\in G_\ell$ on these scalars also affects $\hat{\phi}$:
\begin{equation}
g_\ell\,\mathbb{L}(\phi)=g_\ell\,\mathbb{L}_\ell(\varphi_\ell)\,\hat{\mathbb{L}}(\hat{\phi})=\mathbb{L}_\ell(\varphi'_\ell)\,\hat{\mathbb{L}}(\hat{\phi}')\,h_\ell\,,
\end{equation}
where $h_\ell\in H_\ell$ is the compensating transformation in:
$$g_\ell\,\mathbb{L}_\ell(\varphi_\ell)=\mathbb{L}_\ell(\varphi'_\ell)\,h_\ell\,,$$
and 
$$\hat{\mathbb{L}}(\hat{\phi}')\equiv h_\ell\,\hat{\mathbb{L}}(\hat{\phi})\,h_\ell^{-1}\,.$$
Since, for our analysis, we are interested in a symmetric submanifold $M_0$ of ${M}_{{\rm scal.}}$ at fixed, non-vanishing values $\hat{\phi}_*\equiv (\phi^k_*)$ of $\hat{\phi}=(\phi^k)$ (to be identified with the extremum of the scalar potential) we will have to restrict the moduli fields to a symmetric submanifold $G_0/H_0$
of $G_\ell/H_\ell$ ($G_0\subset G_\ell,\,H_0\subset H_\ell$) characterized by the property that $[G_0,\,\mathbb{L}(\hat{\phi}_*)]=0$.
To this end, we first reduce $\hat{\phi}_*$ to its simplest form (\emph{normal form}), by acting on it using $H_\ell$ and then define $G_0$ to be a non-compact, semisimple subgroup of $G_\ell$ commuting with $\mathbb{L}(\hat{\phi}_*)$. The manifold $M_0=G_0/H_0$,$H_0$ being the maximal compact subgroup of $G_0$, is now a symmetric submanifold of ${M}_{{\rm scal.}}$ at $\hat{\phi}=\hat{\phi}_*$ and, as such, it is \emph{totally geodesic}.\footnote{A totally geodesic submanifold $M_0$ of a Riemannian manifold ${M}_{{\rm scal.}}$ is characterized by the property that a geodesic in ${M}_{{\rm scal.}}$ originating in a point of $M_0$ and initially tangent to $M_0$, lies entirely in the submanifold itself. } Let us denote a point in it by $\varphi_0\equiv (\varphi^a)$, $a=1,\dots,\,n_m$, where $n_m<n_\ell$ is the dimension of $M_0$.
\par
Using the ansatz \eqref{AnsatzH} and the property of $\mathcal{M}$ of being pseudo-orthogonal symmetric,
we can write
\begin{equation}
    \mathbb{H}^T\frac{\partial}{\partial \phi^s} \mathcal{M}\,\mathbb{H}=4\,\frac{\partial}{\partial \phi^s}V(\phi,\Gamma)\,L^{-2d}\,,
\end{equation}
where:
\begin{equation}
   V(\phi,\Gamma)\equiv \frac{1}{2}\,\Gamma^T\mathcal{M}(\phi)\Gamma\,.
\end{equation}
The scalar field equations, on the ansatz, read:
\begin{align}
       D_{\hat{\mu}}(\partial^{\hat{\mu}} \phi^s)= \nabla_{\hat{\mu}} (\partial^{\hat{\mu}} \phi^s)+\tilde{\Gamma}^s_{uv}\partial_{\hat{\mu}} \phi^u \partial^{\hat{\mu}} \phi^v&=\mathcal{G}^{st}\,\partial_t V\,L^{-2d}\, \label{ScalarEqGen2}.
\end{align}
Notice that, in the chosen parametrization defined by the coset representative of the form \eqref{coset01}, the scalar potential is independent of the moduli fields $\varphi^u$. Indeed, by using eqs. \eqref{ML} and \eqref{littleG}:
$$V(\varphi^s,\Gamma^M)=V(\varphi^u,\,\phi^k,\Gamma^M)=V(\phi^k,\,\Gamma^M)\,.$$
Let us define $\hat{\phi}_* \equiv (\phi^k_*)$ to be the value of the non-moduli fields $\phi^k$ which extremize the scalar potential $V$:
\begin{equation}
    \left.\frac{\partial V}{\partial \phi^k}\right\vert_{\phi^k=\phi^k_*}=0\,\,\Rightarrow \,\,\phi^k_*=\phi^k_*(\Gamma)\,.
\end{equation}
The fixed values $\hat{\phi}_*=(\phi^k_*)$ only depend on the quantized charges and are defined modulo the action of a $H_\ell$ transformation. In light of our previous discussion, we fix the action of $H_\ell$ to bring  $\hat{\phi}_*$ to its normal form and then restrict the moduli fields to the coordinates $\varphi^a$, $a=1,\dots, n_m$, of the symmetric submanifold $M_0=G_0/H_0$. 
The value of the scalar potential at the minimum is denoted by $V_*$ and only depends on the quantized charges $\Gamma^M$:
$$V_*(\Gamma)=V(\phi^k_*,\Gamma)>0\,.$$
We complete our ansatz by choosing
\begin{equation}
   \phi^k(x)\equiv\phi^k_* \,\,,\,\,\,\,\varphi^a(x)=\varphi^a(\eta)\,,
\end{equation}
where $\eta$ is the coordinate parametrizing $S^1$ in ${\rm M}_d={\rm AdS}_{d-1}\times S^1$. All other scalar fields are set to zero.
The scalar field equation reduces to the geodesic equation for $\varphi^a(\eta)$ on ${M}_0$:
\begin{equation}
\ddot{\varphi^a}+\tilde{\Gamma}^a{}_{bc}\,\dot{\varphi}^b\,\dot{\varphi}^c=g_{\eta\eta}\,\mathcal{G}^{ak}\,\left.\frac{\partial V}{\partial \phi^k}\right\vert_{\phi^k=\phi^k_*}\,L^{-2d}=0\,,
\end{equation}
where we have used the fact that $V$ is independent of $\varphi^a$ and have defined $\dot{\varphi}\equiv d\varphi/d\eta$.
Having assumed the non-moduli fields $\phi^k$ to be constant and equal to the values that minimize $V$, all the derivatives of $\phi^k$ on the solution vanish and the corresponding field equations read:
$$g^{\eta\eta}\,\tilde{\Gamma}^k{}_{bc}\,\dot{\varphi}^b\,\dot{\varphi}^c=L^{-2d}\,\mathcal{G}^{kk'}\,\left.\frac{\partial V}{\partial \phi^{k'}}\right\vert_{\phi^k=\phi^k_*}\,L^{-2d}=0\,,$$
which is satisfied provided $\tilde{\Gamma}^k{}_{bc}=0$ for $\phi^k=\phi^k_*$ and $\varphi^a$ generic. This condition follows from the property of the moduli space $G_0/H_0$ of being a totally geodesic submanifold of $G/H$.\footnote{Notice indeed that the coset space $\mathbb{K}_0\subset \mathbb{K}$ of $G_0/H_0$ is also contained in the tangent space $\mathbb{K}_*\equiv \hat{\mathbb{L}}(\phi_*^k)\,\mathbb{K}\,\hat{\mathbb{L}}(\phi_*^k)^{-1}$ to $M_{{\rm scal.}}$ at the point $\varphi^u=0,\,\phi^k=\phi^k_*$ since $\mathbb{K}_0=\hat{\mathbb{L}}(\phi_*^k)\,\mathbb{K}_0\,\hat{\mathbb{L}}(\phi_*^k)^{-1}\subset \mathbb{K}_*$. Being $M_0$ symmetric, it follows that $[[\mathbb{K}_0,\,\mathbb{K}_0],\,\mathbb{K}_0]\subset \mathbb{K}_0$, namely $\mathbb{K}_0$ is a \emph{Lie triple system} and thus $M_0$ is totally geodesic \cite{Helgason}.}\footnote{By the same token, also the field equations for the moduli fields $\varphi^u$, $u\neq a$, which are set to zero, are satisfied since $\tilde{\Gamma}^u{}_{bc}=0$, $u\neq a$.}\\
The moduli fields $\varphi^a(\eta)$, therefore, describe a geodesic in $M_0$. Let us denote by $\kappa$ the corresponding line element (or "velocity") along it:
\begin{equation}
    \kappa^2=\frac{1}{2}\,\mathcal{G}_{ab}\,\dot{\varphi}^a\,\dot{\varphi}^b\,.
\end{equation}
The reader can verify that the tensor-field equations $d\mathbb{H}^M=0$ are satisfied by the ansatz.\par As far as the Einstein equations are concerned, the only non-vanishing components of the Riemann and Ricci tensors are:
\begin{align}
    R_{\alpha\beta\gamma\delta}&= -v_1^{-2}\,(g_{\alpha \gamma} g_{\beta \delta}-g_{\alpha \delta} g_{\beta \gamma})\,\,\Rightarrow,\,\,\,R_{\alpha\beta}=R_{\alpha\gamma\beta}{}^\gamma=-\frac{(d-2)}{v_1^2}\,g_{\alpha\beta}\,,\nonumber\\
   R_{ijkl}&= v_2^{-2}\,(g_{ik} g_{jl}-g_{il} g_{jk})\,\,\Rightarrow,\,\,\,R_{ij}=R_{ikj}{}^k=\frac{(d-1)}{v_2^2}\,g_{ij}\,.  
\end{align}
The Einstein equation can be conveniently recast in the form:
\begin{equation}
R_{\hat{\mu}\hat{\nu}}=\,\frac{1}{2}\mathcal{G}_{rs}\,\partial_{\hat{\mu}} \phi^r \partial_{\hat{\nu}}\phi ^s+T^{(H)}_{\hat{\mu}\hat{\nu}}\,.
\end{equation}
The components of $T^{(H)}$ are:
\begin{align} T^{(H)}_{\alpha\beta}&=-\frac{1}{2}\,V_*\,L^{-2d}\,g_{\alpha\beta}\,,\nonumber\\
T^{(H)}_{\eta\eta}&=-\frac{1}{2}\,V_*\,v_3^2\,L^{-2d}\,,\nonumber\\
T^{(H)}_{ij}&=\frac{1}{2}\,V_*\,L^{-2d}\,g_{ij}\,.
\end{align}
Taking into account the contribution to the energy-momentum tensor coming from the geodesic motion of the moduli fields $\varphi^a$, the Einstein equations imply the following relations:
\begin{equation}
    v_1^2=\frac{2(d-2)\,L^{2d}}{V_*}\,\,,\,\,\, v_2^2=L^2=\frac{2(d-1)\,L^{2d}}{V_*}\,,\,\,\,v_3^2=\frac{2\,\kappa^2\,L^{2d}}{V_*}\,.
\end{equation}
From the second equation, it follows that:
$$L^{2(d-1)}=\frac{V_*}{2(d-1)}\,,$$
and, in terms of $L$, the radii $v_1$ of ${\rm AdS}_{d-1}$ and $v_3$ of $S^1$ read
$$v_1=\sqrt{\frac{d-2}{d-1}}\,L\,,\,\,\,v_3=\,\frac{\kappa}{\sqrt{d-1}}\,L\,.$$
We notice that the radius $v_3$ of $S^1$ is proportional to the "velocity" along the geodesic, that is its constant unit measure $\kappa$. \\
This solution is to be compared to the known one
with geometry ${\rm AdS}_d\times S^d$
in which 
$$
\phi^k=\phi^k_*(\Gamma)={\rm const.}\,\,,\,\,\,\varphi^u={\rm const.}\,.
$$
In that case, the ansatz for the metric reads
\begin{equation}
ds^2=g_{\mu\nu}\,dx^\mu\,dx^\nu+g_{ij}\,d\upxi^i\,d\upxi^j=v_1^{2}\,ds^2_{{\rm AdS}_{d}}+v_2^{2}\,ds^2_{S^d}\,,
\end{equation}
and that for the tensor field strengths has the same general form \eqref{AnsatzH}.
For this solution, we find:
\begin{equation}
    v_1^2=v_2^2=L^2=\frac{2(d-1)\,L^{2d}}{V_*}\,\,\Rightarrow\,\,\,v_1=L=\left(\frac{V_*}{2(d-1)}\right)^{\frac{1}{2(d-1)}}\,.\label{adsdsdcase}
\end{equation}
Notice that the ${\rm AdS}_{d-1}\times S^1\times S^d$ discussed above and the known ${\rm AdS}_d\times S^d$ solution are not continuously connected. Indeed setting $\kappa=0$ amounts in the former to shrinking $S^1$ to zero radius. We may say that the ${\rm AdS}_{d-1}\times S^1\times S^d$ solution may be obtained from the ${\rm AdS}_d\times S^d$ by compactifying one direction on the boundary of ${\rm AdS}_d$ and giving the moduli fields $\varphi^a$ a geodesic dependence on the 
coordinate of the corresponding $S^1$. This procedure amounts to performing a Scherk-Schwarz reduction on $S^1$ with a twist matrix described by a \emph{hyperbolic element} of $G_0$, as we will show below. It is the back-reaction of the evolving scalar fields on space-time that deforms ${\rm AdS}_d$ into ${\rm AdS}_{d-1}\times S^1$.

It is interesting to note that this construction can be generalized. We can take $k$-directions along the boundary of  AdS$_{d}$ and compactify them, allowing the moduli to depend on the coordinates $\eta_1,\dots, \eta_k$ along these $k$ directions:
$$\varphi^a=\varphi^a(\boldsymbol{\eta})=\varphi^a(\eta_1,\dots, \eta_k)\,,$$
where $\boldsymbol{\eta}\equiv (\eta_1,\dots, \eta_k)$, while the non-moduli fields $\phi^k$ are still fixed at their attractor values $\phi^k_*$. We assume $d-k>1$.
Let us further assume that there is a symmetric submanifold of $G_\ell/H_\ell$ consisting of the product of $k$-factors:
\begin{equation}
M_0=\frac{G_0}{H_0}=\frac{G_1}{H_1}\times \frac{G_2}{H_2}\times \dots \times\frac{G_k}{H_k}\,,
\end{equation}
where the elements of $G_0=G_1\times \dots \times G_k$ are required to commute with $\hat{\mathbb{L}}(\hat{\phi}_*)$. Then we take $\varphi^a(\eta_1,\dots, \eta_k)$ to describe $k$ geodesics, one within each factor $\frac{G_{\tt j}}{H_{\tt j}}$, ${\tt j}=1,\dots, k$,  so that:
$$\frac{1}{2}\,\mathcal{G}_{ab}(\varphi(\boldsymbol{\eta}))\partial_{\eta_{\tt i}}\varphi^a\partial_{\eta_{\tt j}}\varphi^b=\delta_{\tt ij}\kappa^2_{\tt i}\,.$$
The back-reaction of the moduli on the background yields a spacetime with geometry:
\begin{align}
    {\rm AdS}_{d-k}\times \underbrace{S^1\times \dots \times S^1}_k\times S^d
    \,.\label{AdS2SSSSSd}
\end{align}
Writing the metric in the form
\begin{equation}
ds^2=g_{\mu\nu}\,dx^\mu\,dx^\nu+g_{ij}\,d\upxi^i\,d\upxi^j=v_1^{2}\,ds^2_{{\rm AdS}_{d-k}}+\sum_{\tt j=1}^k\,v_{2+\tt j}^{2}\,d\eta_{\tt j}^2+v_2^{2}\,ds^2_{S^d}\,,
\end{equation}
where $\eta_1,\dots, \eta_k$ parameterize the $k$ circles, we find, from the Einstein equation:
\begin{equation}
    v_1^2=\frac{2(d-k-1)\,L^{2d}}{V_*}\,\,,\,\,\, v_2^2=L^2=\frac{2(d-1)\,L^{2d}}{V_*}\,,\,\,\,v_{\tt j+2}^2=\frac{2\,\kappa_{\tt j}^2\,L^{2d}}{V_*}\,,\,\,{\tt j}=1,\dots k\,.
\end{equation}
Having assumed $d-k>1$, $v_1$ is non-vanishing.
The above equations imply
$$L^{2(d-1)}=\frac{V_*}{2(d-1)}\,,\,\,v_1=\sqrt{\frac{d-1-k}{d-1}}\,L\,,\,\,\,v_{\tt j+2}=\,\frac{\kappa_{\tt j}}{\sqrt{d-1}}\,L\,.$$
In this section, we proved that the general ansatz describing the new solutions satisfies the field equations. In this, no role is played by the global properties of the solutions themselves. In particular, we could have also considered our backgrounds in the form 
\begin{align}
    {\rm AdS}_{d-k}\times \underbrace{\mathbb{R}\times \dots \times \mathbb{R}}_k\times S^d \,.
\end{align}
For $d=5$ and $k=1$ the background is a singular instance of Janus solutions \cite{Bak:2003jk,Clark:2004sb,DHoker:2007zhm,DHoker:2007hhe} in Type IIB superstring theory.\par 
However, since the geodesics on the moduli space $G_0/H_0$ are non-compact, if the affine parameter $\eta_j$ varies in $\mathbb{R}$, the corresponding geodesic would stretch to the boundary of the manifold and the solution would not be regular.  Considering the product of $k$ circles $S^1$, we need to specify, in a consistent way, the corresponding boundary conditions of the bosonic fields.\footnote{Alternatively, we could consider a product of $k$ finite intervals (${I}^k$) or of $k$ circles.} We shall expand on this issue, within superstring theory, in the next subsection.\par
Let us conclude this section by noting that if we started from the background ${\rm EAdS}_d\times S^d$ in the Euclidean version of the model, the moduli space 
would be a pseudo-Riemannian, Wick-rotated version of the one in the Lorentzian theory. As such it can also describe time-like geodesics for which
\begin{equation}
    \frac{1}{2}\,\mathcal{G}_{ab}\,\dot{\varphi}^a\,\dot{\varphi}^b=-\kappa^2<0\,.
\end{equation}
These geodesics would allow for a decomposition of the $d$-sphere yielding geometries of the form:
\begin{align}
    {\rm EAdS}_{d}\times \underbrace{S^1\times \dots \times S^1}_k\times S^{d-k}
    \,.\label{EAdS2SSSSSd}
\end{align}
\subsection{Boundary conditions on $S^1$}\label{bc}
Consider the solution with spacetime of the form $ {\rm AdS}_{d-1}\times S^1\times S^d$ within superstring theory. Let us denote by $T$ the length of $S^1$ so that $\eta\in [0,\,T)$. Since the dependence of the moduli fields $\varphi^a(\eta)$ on $\eta$ describes a geodesic on $M_0$ which connects two distinct points as we move around the circle $S^1$, consistency of the background as a string theory solution requires the \emph{monodromy} matrix connecting the initial and final points to be a string duality, namely to belong to $G_0(\mathbb{Z})$. 
More precisely, suppose $\varphi^a(\eta)$ describe a geodesic on $M_0$ connecting an initial point, which we can choose to be the origin of the moduli space $\varphi^a(0)=0$, to a final point in $M_0$ with coordinate $\varphi^a(T)$. The coset representative along the geodesic defines an $\eta$-dependent "twist" matrix in $G_0/H_0$:
\begin{equation}
    \mathcal{A}(\eta)\equiv \mathbb{L}_0(\varphi^a(\eta))\,.
\end{equation}
The monodromy matrix reads:
\begin{equation}
    \mathfrak{M}\equiv  \mathcal{A}(T)\, \mathcal{A}(0)^{-1}\,,\label{MonodromyM}
\end{equation}
and maps the coset representative in $M_{{\rm scal.}}$ at $\eta=0$ to that at $\eta=T$:
\begin{equation}
     \mathfrak{M}\,\mathbb{L}(\varphi^a(0),\,\hat{\phi}_*)=\mathbb{L}(\varphi^a(T),\,\hat{\phi}_*)\,.
\end{equation}
This matrix is defined modulo a compensating transformation in the isotropy group of $\varphi^a(0)$ in $M_0$:
\begin{equation}
     \mathfrak{M}\,\sim \, \mathfrak{M}\, h_0\,\,,\,\,\,h_0\in \mathbb{L}_0(\varphi^a(0))\,H_0\,\mathbb{L}_0(\varphi^a(0))^{-1}\,.
\end{equation}
We require one representative of the equivalence class of $\mathfrak{M}$ to be an integer matrix:
\begin{equation}
    \mathfrak{M}\in G_0(\mathbb{Z})\,.
\end{equation}
A conjugation of $ \mathfrak{M}$ by an element of $G_0(\mathbb{Z})$ can be reabsorbed by a redefinition of the scalar fields $\varphi^a$ in the background so that it is unphysical. The monodromy matrix is therefore defined by conjugacy classes of hyperbolic elements in $G_0(\mathbb{Z})$.\par
The geodesic on $M_0$ is uniquely defined, by the initial point $\varphi_0(0)\equiv (\varphi^a(0))$ and by a velocity vector 
$\mathbb{Q}\in T_{\varphi_0(0)}(M_0)$. The latter can be described by a matrix of the form
$$\mathbb{Q}=\mathbb{L}_0(\varphi_0(0))\,\mathbb{Q}_0\,\mathbb{L}_0(\varphi_0(0))^{-1}\,,$$
where $\mathbb{Q}_0$, being an element fo the coset space of $M_0$, in a suitable basis of the chosen real matrix representation, is a symmetric matrix: $\mathbb{Q}_0=\mathbb{Q}_0^T$.
The moduli along the corresponding geodesic $\varphi_0(\eta)\equiv (\varphi^a(\eta))$ are solutions to the matrix equation:
\begin{equation}
    \mathcal{M}_0(\varphi_0(\eta))\equiv 
    \mathbb{L}_0(\varphi_0(\eta)) \mathbb{L}_0(\varphi_0(\eta))^T= \mathcal{M}_0(\varphi_0(0))\,e^{\mathbb{Q}^T\,\eta}=\mathbb{L}_0(\varphi_0(0))\,e^{\mathbb{Q}_0\,\eta}\,\mathbb{L}_0(\varphi_0(0))^T\,.
\end{equation}
In terms of $\mathbb{Q}_0$ the "velocity" $\kappa$ along the geodesic reads:
\begin{equation}
\kappa^2=\frac{{\tt k}}{8}\,{\rm Tr}(\mathbb{Q}_0^2)\,.
\end{equation}
We can restrict ourselves to geodesics originating in the origin of $M_0$, $\varphi^a(0)=0$, where $\mathbb{L}_0(\varphi_0(0))=\mathbb{L}_0(0)={\bf 1}$. Having chosen $\mathfrak{M}\in G_0(\mathbb{Z})$, then $T$ and $\mathbb{Q}_0$ have to be fixed so that:
\begin{equation}
 \mathcal{M}_0(\varphi_0(T))=\mathfrak{M}\, \mathcal{M}_0(\varphi_0(0))\,\mathfrak{M}^T=\mathfrak{M}\,\mathfrak{M}^T=e^{\mathbb{Q}_0\,T}\,.\label{georigin}
\end{equation}
In the more general solution (\ref{AdS2SSSSSd}), we have a monodromy matrix $\mathfrak{M}_{\tt j}$ associated with each of the $k$ 1-cycles. Consistency as a string background requires, for each of these matrices:
\begin{equation}
\mathfrak{M}_{\tt j}\in G_{\tt j}(\mathbb{Z})\subset G(\mathbb{Z})\,\,,\,\,\,\,{\tt j}=1,\dots, k\,.
\end{equation}
The solutions described here are instances of $U$-folds \cite{Dabholkar:2002sy}. They feature non-contractible 1-cycles along which there is a monodromy matrix is a string duality in $G(\mathbb{Z})$. In general, this duality is a combination of $S$ and $T$-dualities and it is referred to as $U$-duality.
\subsubsection{Deformations of the global geometry of the solution}\label{chis}
Let us discuss a deformation of the global geometry of the solution effected by introducing suitable metric moduli. Since $d$ is an odd number, it is convenient to choose the following  parametrization for $S^d$ 
\begin{equation}
   \{ \upxi^i\}=\{\mu_I,\,\upphi_I\}\,,\,\,\,i=1,\dots, d\,,\,\,\,I=1,\dots, {\tt p}=\frac{d+1}{2}\,.
\end{equation}
where $\mu_I\ge 0,\,\upphi_I\in[0,\,2\pi)$, 
in which its metric reads:
\begin{equation}
    ds^2_{S^d}=L^2\sum_{I=1}^{\tt p}(d\mu_I^2+\mu_I^2\,d\upphi_I^2)\,.
\end{equation}
The coordinates $\mu^I$ are subject to the constraint:
$\sum_{I=1}^{\tt p}\mu_I^2=1\,.$
Let us now perform, on the solution discussed in the previous sections, the following local change of variables:
$$\upphi_I\,\rightarrow\,\upphi_I'=\upphi_I+\chi_I\,\eta\,.$$
The resulting solution is locally, though not globally equivalent to the original one, as discussed in \cite{Giambrone:2021zvp,Guarino:2021kyp,Guarino:2021hrc,Guarino:2022tlw}.
The parameters $\chi_I$ are metric deformation of the manifold $S^1\times S^d$ which can be understood as follows.
Consider the \emph{local} torus $$T^{{\tt p}+1}=S^1_{\upphi_1}\times S^1_{\upphi_2}\times \dots \times S^1_\eta\,,$$
where $S^1_{\upphi_I}$ are the local circles parametrized by $\upphi_I$ and $S^1_\eta$ is the circle parametrized by $\eta$.
For $\chi_I=0$, $T^{{\tt p}+1}$  can be described as the quotient
$$T^{{\tt p}+1}=\mathbb{R}^{{\tt p}+1}/\Lambda_{{\rm cubic}}\,,$$
where $\Lambda_{{\rm cubic}}$ is a cubic lattice generated by an orthonormal basis of vectors ${\bf u}_I,\,{\bf u}$. 
The position vector ${\bf x}$ of a point in $T^{{\tt p}+1}$ is defined modulo a vector in the lattice  (summation over the repeated index $I$ is understood):
\begin{equation}
    {\bf x}= \frac{\upphi_I}{2\pi}\,{\bf u}_I+\frac{\eta}{T}\,{\bf u}\sim {\bf x}+n_I\,{\bf u}_I+n\,{\bf u}\,\,;\,\,\,n_I,\,n\in \mathbb{Z}\,,
\end{equation}
which implies $\upphi_I\sim \upphi_I+ 2\pi$ and $\eta\sim \eta+T $.\par
After the deformation, that is for $\chi_I\neq 0$, the local torus 
is described as:
$$T^{{\tt p}+1}=\mathbb{R}^{{\tt p}+1}/\Lambda[\chi_I]\,,$$
where $\Lambda[\chi_I]$ is a lattice generated by the non-orthogonal basis $\{\tilde{{\bf u}}_I,\,\tilde{\bf u}\}$, where
$$\tilde{{\bf u}}_I\equiv {{\bf u}}_I\,\,,\,\,\,\,\tilde{\bf u}\equiv \frac{\chi_I T}{2\pi}\,{\bf u}_I+{\bf u}\,.$$
Now the position vector of a point on the local torus is subject to the identification:
\begin{equation}
    {\bf x}= \frac{\upphi_I}{2\pi}\,{\bf u}_I+\frac{\eta}{T}\,\tilde{{\bf u}}\sim {\bf x}+n_I\,{\bf u}_I+n\,\tilde{{\bf u}}\,\,,\,\,\,n_i,\,n\in \mathbb{Z}\,,
\end{equation}
which implies the same identifications on $\upphi_i,\,\eta$ given earlier. Defining instead new coordinates $\upphi'_I,\,\eta'$ with respect to the old orthonormal basis $${\bf x}= \frac{\upphi'_I}{2\pi}\,{\bf u}_I+\frac{\eta'}{T}\,{\bf u}\,,$$
we have the following relation:
$$\upphi_I'=\upphi_I+\chi_I\,\eta\,,\,\,\eta'=\eta\,.$$
While in the coordinates $\{\upphi_I,\,\eta\}$ the parameters $\chi_I$ appear in the metric, in the coordinates   $\{\upphi'_I,\,\eta\}$, they only appear in \emph{twisted} identifications:
\begin{equation}
    \upphi'_I\sim  \upphi'_I+2\pi\,\,,\,\,\,\eta\sim \eta+T \,\wedge \,\,  \upphi'_I\sim  \upphi'_I+\chi_I\,T\,.
\end{equation}
The deformations $\chi_I$ break the isometry group ${\rm SO}(d+1)$ of $S^d$ to its maximal torus ${\rm U}(1)^{\tt p}$ acting on $\upphi_I$ as shift-isometries. \par This global deformation was originally studied in \cite{Giambrone:2021zvp,Guarino:2021kyp,Guarino:2021hrc,Guarino:2022tlw} in 
more sophisticated variants of the U-folds described here, for $d=5$, and the parameters $\chi_I$ corresponded to exactly marginal deformations of the dual SCFT at the boundary.\par
From the above characterization, it follows that $\chi_I$ are periodic since:
\begin{equation}
    \chi_I\rightarrow \chi_I\pm \frac{2\pi}{T}\,\,\Rightarrow\,\,\,\tilde{{\bf u}}_I\,\rightarrow \, \tilde{{\bf u}}_I \,\,,\,\,\,\tilde{{\bf u}}\,\rightarrow\, \tilde{{\bf u}}\pm\tilde{{\bf u}}_I\in \Lambda[\chi_I]\,,
\end{equation}
so that $\chi_I\sim \chi_I\pm  \frac{2\pi}{T}$.
\subsection{Examples in Type IIB superstring theory}\label{examples}
\paragraph{Type IIB in $D=10$.}
The first setting where our construction can be applied is Type IIB theory in ten dimensions  \cite{Schwarz:1983qr}. The global symmetry group of the classical low-energy supergravity 
description is $G={\rm SL}(2,\mathbb{R})_{{\rm IIB}}$ within which only $G(\mathbb{Z})={\rm SL}(2,\mathbb{Z})_{{\rm IIB}}$ is a symmetry of the superstring theory.
It is known that this theory admits maximally supersymmetric background of the form ${\rm AdS}_5\times S^5$ in which the 3-form field strengths vanish and the self-dual five-form field strength $\hat{F}_{(5)}={}^*\hat{F}_{(5)}$ has a non-vanishing flux, related to the radii of the ${\rm AdS}_5$ and $S^5$. 
The construction discussed earlier applies with $d=5,\,p=4$. In this case the representation $\mathscr{R}_p=\mathscr{R}_4={\bf 1}$, namely it is the singlet representation since the electric and magnetic 5-form field strengths coincide $G=F=\mathbb{H}=\hat{F}_{(5)}$, $\mathcal{M}(\phi)=1$. Moreover, $G_0=G$ and the scalar manifold 
$$M_{{\rm scal.}}=\frac{G}{H}=\frac{{\rm SL}(2,\mathbb{R})_{{\rm IIB}}}{{\rm SO}(2)}\,,$$
coincides with the moduli space $M_0=G_0/H_0$ and describes the axion $C_{(0)}$ and the dilaton $\phi$. In our previous notation we identify $\varphi^a=(\phi,\,C_{(0)})$ and the (left) coset representative in the ${\bf 2}$ of ${\rm SL}(2,\mathbb{R})$ is $\mathbb{L}=\mathbb{L}_0(\varphi^a)=L_{(2)}(\phi,\,C_{(0)})$, where $L_{(2)}$ is defined in eq. \eqref{L2}
 below.\footnote{ In this representation, the action from the left of a matrix $\left(\begin{matrix}a & b\cr c & d \end{matrix}\right)$, with $ad-bc=1$, amounts to the following transformation on the complex axion-dilaton field $\rho=C_{(0)}+i\,e^{-\phi}$: $\rho\rightarrow \rho'=(d \rho+c)/(b\rho+a)$. }
The background with geometry ${\rm AdS}_4\times S^1\times S^5$ was constructed in \cite{Guarino:2019oct}. The axion and the dilaton describe a geodesic on $M_{{\rm scal.}}$. This background is non-supersymmetric and unstable.
The solution in \cite{Guarino:2019oct} can be generalized by considering a geodesic in the moduli space originating in a generic point $\phi(0),\,C_{(0)}(0)$ with "velocity" $\kappa$:
\begin{equation}
    e^{\phi(\eta)}=\cosh(\sqrt{2} \kappa\,\eta)\, e^{\phi(0)}\,\,,\,\,
   C_{(0)}(\eta)=  e^{-\phi(0)}\tanh \left(\sqrt{2} \kappa\,\eta   \right)+C_{(0)}(0)\,.
\end{equation}
This solution, however, does not satisfy the appropriate boundary conditions for it to be a possible solution to Type IIB superstring theory: the monodromy matrix as $\eta\rightarrow \eta+T$ must be in ${\rm SL}(2,\mathbb{Z})_{{\rm IIB}}$. Choosing, for instance, \footnote{In the definition of $J_{\tt n}$,  $\mathcal{S}\equiv \left(\begin{matrix}0 & -1\cr 1 & 0\end{matrix}\right)$ and $\mathcal{T}\equiv \left(\begin{matrix}1 & 0\cr 1 & 1\end{matrix}\right)$. $J_{\tt n}$ is an elliptic element of ${\rm SL}(2,\mathbb{Z})$ for ${\tt n}=0,1$, parabolic for ${\tt n}=2$ and hyperbolic for ${\tt n}>2$. It was shown that hyperbolic elements of ${\rm SL}(2,\mathbb{Z})$, modulo conjugations, can be either brought to the standard form $J_{\tt n}$, ${\tt n}>2$,
or to coincide with sporadic monodromies, see \cite{Dabholkar:2002sy}. }
\begin{equation}
    \mathfrak{M}=J_{\tt n}\equiv -\mathcal{S}\,\mathcal{T}^{\tt n}=\left(\begin{matrix} {\tt n} & 1\cr -1 & 0\end{matrix}\right)\,\,,\,\,{\tt n}\in \mathbb{Z}\,,\,\,{\tt n}>0\,,
\end{equation}
and $C_{(0)}=\phi(0)=0$, solving eq. \eqref{georigin}, we find the following solution:
{\small \begin{align}
    e^{\phi(\eta)}&=\frac{{\tt n} \sinh \left(\sqrt{2}\kappa \eta\right)}{\sqrt{{\tt n}^2+4}}+\cosh \left(\sqrt{2}\kappa \eta\right)\,\,,\,\,
   C_{(0)}(\eta)=-\frac{2 \sinh \left(\sqrt{2}\kappa \eta\right)}{\sqrt{{\tt n}^2+4} \cosh \left(\sqrt{2}\kappa \eta\right)+{\tt n}\sinh \left(\sqrt{2}\kappa \eta\right)}\,,\nonumber\\
   T&=\frac{1}{\sqrt{2} \kappa }\cosh ^{-1}\left(\frac{{\tt n}^2}{2}+1 \right)\,.
\end{align}}
Defining the complex field $\rho=C_{(0)}+i\,e^{-\phi}$ one can verify that 
\begin{equation}
    \rho(T)=-\frac{1}{\rho(0)+{\tt n}}\,,
\end{equation}
which is the effect of the transformation $J_{\tt n}=-\mathcal{S}\,\mathcal{T}^{\tt n}$ on $\rho(0)$.
Notice that, on this solution 
$$\mathfrak{M}=\mathcal{A}(T)\,\mathcal{A}(0)^{-1}\,h_0=L_{(2)}(\phi(T),\,C_{(0)}(T))L_{(2)}(\phi(0),\,C_{(0)}(0))^{-1}\,h_0\,,$$
where the compensating transformation $h_0$ reads:
\begin{equation}
    h_0=\frac{1}{\sqrt{1+{\tt n}^2}}{}\left(\begin{matrix} {\tt n} & 1\cr -1 & {\tt n}\end{matrix}\right)\in {\rm SO}(2)\,.
\end{equation}

We could start from the background of the form \cite{Kachru:1998ys,Corrado:2002wx,Louis:2015dca}:
\begin{equation}
    {\rm AdS}_5\times S^5/\mathbb{Z}_k\,,
\end{equation}
which preserves 16 supercharges. The classical supergravity description features the following moduli space
\begin{equation}
M_{{\rm scal.}}=\frac{{\rm SU}(1,k)}{{\rm U}(1)\times {\rm SU}(k)}\,.    
\end{equation}
It describes $k$ complex scalars dual to the complexified coupling constants $\tau_i$ of the dual $\mathcal{N}=2$ supersymmetric $A_{k-1}$ quiver gauge theory. Also in this case $\mathscr{R}_p=\mathscr{R}_4={\bf 1}$, $G=F=\mathbb{H}=\hat{F}_{(5)}$, $\mathcal{M}(\phi)=1$ and $G_0=G={\rm SU}(1,k)$. 
Applying our construction to this situation would require the monodromy matrix to belong to the quantum duality which generalizes the ${\rm SL}(2,\mathbb{Z})$ symmetry for the $k=1$ case to $k>1$. For a discussion on the definition of this group and its relation to the classical symmetry ${\rm SU}(1,k)$ of the supergravity moduli space see, for instance, \cite{Dorey:2001qj,Corrado:2002wx}.
\paragraph{Type IIB on $T^4$ or $K3$}
The six-dimensional theory resulting from the compactification of Type IIB superstring theory on a 4-torus $T^4$ is described, in its low-energy limit, by the maximal $\mathcal{N}=(2,2)$ six-dimensional supergravity \cite{Tanii:1984zk}.
This theory features a solution with spacetime ${\rm AdS}_3\times S^3$, describing the near-horizon geometry of a system of D1-D5 branes with two common Neumann directions (at the boundary of ${\rm AdS}_3$) and the five-branes wrapping the 4-torus. Alternatively, in an S-dual picture, the same background describes the near-horizon geometry of an F1-NS5 system.\par The global symmetry group $G$ of the classical six-dimensional supergravity is ${\rm Spin}(5,5)$, double cover of ${\rm SO}(5,5)$. Quantum effects break this group to $G(\mathbb{Z})={\rm Spin}(5,5;\,\mathbb{Z})$, which is conjectured to encode superstring U-dualities. This group acts on the charge lattice $\boldsymbol{\Gamma}^{5,5}$. The bosonic sector of the theory describes $n=5$ 2-forms, $25$ scalar fields in the coset (in the classical theory) 
$$M_{{\rm scal.}}=\frac{G}{H}=\frac{{\rm Spin}(5,5)}{[({\rm Spin}(5)\times {\rm Spin}(5))/\mathbb{Z}_2]}\,,$$
and 16 vector fields in the ${\bf 16} $ of ${\rm Spin}(5,5)$. The ten self-dual and anti-self-dual components of the 3-form field strengths $\mathbb{H}^M_{\hat{\mu}\hat{\nu}\hat{\rho}}$, $M=1\dots, 10$, transform, under the action of the classical global symmetry group ${\rm SO}(5,5)$ in the representation $\mathscr{R}_2={\bf 10}$. On the bosonic backgrounds under consideration, the vector fields vanish. Since we shall be focusing on the bosonic sector only, the global symmetry group $G$ will be described as ${\rm SO}(5,5)$. The charge vector $\Gamma^M$ of the D1-D5 system has little group $G_\ell={\rm SO}(4,5)\subset G$ in the classical supergravity.
The coset space $\mathbb{K}$ of $M_{{\rm scal.}}$, under the adjoint action of $H_\ell={\rm SO}(4)\times {\rm SO}(5)$ splits into $\mathbb{K}_\ell\oplus \hat{\mathbb{K}}$ in the representation ${\bf (4,5)}\oplus {\bf (1,5)}$, see eq. (\ref{KKK}). The non-moduli scalar fields $\hat{\phi}^k$ are chosen to be parameters of $\hat{\mathbb{K}}$ in the ${\bf (1,5)}$ of $H_\ell$. As discussed earlier, we fix ${\rm SO}(5)$ to rotate $\hat{\phi}^k$ to their simplest form. This corresponds to the parameter of a Cartan generator in the coset space. Next, we consider a group $G_0$ commuting with the corresponding element $\hat{\mathbb{L}}(\hat{\phi}^k)$. Our choice for $G_0$ needs not to be maximal. The maximal choice of $G_0$ is ${\rm SO}(4,4)$ and thus of the moduli space $M_0=G_0/H_0={\rm SO}(4,4)/{\rm SO}(4)\times {\rm SO}(4)$.\par
The D5-branes of the D1-D5 system wrap the whole $T^4$ and the D1,\,D5 charges $d_1,\,d_5$ belong to a $\boldsymbol{\Gamma}^{1,1}$ sublattice of $\boldsymbol{\Gamma}^{5,5}$, which thus decomposes as \cite{Seiberg:1999xz}
$$\boldsymbol{\Gamma}^{5,5}=\boldsymbol{\Gamma}^{1,1}\oplus \boldsymbol{\Gamma}^{4,4}\,.$$
The subgroup $G_0$ acts on $\boldsymbol{\Gamma}^{4,4}$ leaving $\boldsymbol{\Gamma}^{1,1}$ invariant. \par
Choosing one direction along ${\rm AdS}_3$ to be compact and fixing the dependence $\varphi^a(\eta)$ of the moduli fields $\varphi^a$ in the space $G_0/H_0$ on the compact boundary coordinate $\eta$, to describe a geodesic in $G_0/H_0$, the back-reaction of these moduli deforms ${\rm AdS}_3\times S^3$ into ${\rm AdS}_2\times S^1\times S^3$ as discussed earlier.
Consistency as a string background requires the monodromy matrix $\mathfrak{M}$, connecting the starting and ending points of the geodesic, to be a \emph{hyperbolic} element of ${\rm SO}(4,4;\,\mathbb{Z})$:
\begin{equation}
   \mathfrak{M}\equiv \mathbb{L}_0(\varphi^a(T))\, \mathbb{L}_0(\varphi^a(0))^{-1}\in {\rm SO}(4,4;\,\mathbb{Z})\,.
\end{equation}
We could also choose the geodesic describing the moduli, to lie in a non-maximal symmetric subspace of $M_{{\rm scal.}}$ such as:
$$M_0=\frac{G_0}{H_0}=\left(\frac{{\rm SL}(2,\mathbb{R})}{{\rm SO}(2)}\right)^4\subset \frac{{\rm SO}(4,4)}{{\rm SO}(4)\times {\rm SO}(4)}\subset  M_{{\rm scal.}}\,.$$
In this particular case, $\mathfrak{M}$ has the general form:
\begin{equation}
\mathfrak{M}=\mathfrak{M}_1\,\mathfrak{M}_2\,\mathfrak{M}_3\,\mathfrak{M}_4\,,
\end{equation}
where $\mathfrak{M}_{\tt j}$, ${\tt j}=1,\dots,4$, is an element of the ${\rm SL}(2,\mathbb{Z})$ subgroup of the corresponding ${\rm SL}(2,\mathbb{R})$ factor in $G_0$. 
We shall expand on this class of U-fold solutions in the maximal six-dimensional theory in the next section.\par
Considering Type IIB superstring theory compactified on $K3$, this is a six-dimensional $(2,0)$ ungauged supergravity \cite{Aspinwall:1996mn}. The group $G$ is ${O}(5,21)$ and the charge lattice is $\boldsymbol{\Gamma}^{5,21}$. Considering a D1-D5-brane system in which the D5 branes wrap the whole $K3$, the little group $G_\ell$ of corresponding charge vector $\Gamma^M$ is ${\rm O}(4,21)$, yielding 84 moduli fields in $G_\ell/H_\ell={\rm SO}(4,21)/{\rm SO}(4)\times {\rm SO}(21)$. According to our construction, we do not consider geodesics within this manifold, but rather within a smaller one $M_0=G_0/H_0$. The maximal choice of $G_0$ is  ${O}(4,20)$. The D1, D5 charges $d_1,\,d_5$ belong to a $\boldsymbol{\Gamma}^{1,1}$ sublattice of $\boldsymbol{\Gamma}^{5,21}$, which thus decomposes as \cite{Seiberg:1999xz}
$$\boldsymbol{\Gamma}^{5,21}=\boldsymbol{\Gamma}^{1,1}\oplus \boldsymbol{\Gamma}^{4,20}\,.$$
The subgroup $G_0$ acts on $\boldsymbol{\Gamma}^{4,20}$ leaving $\boldsymbol{\Gamma}^{1,1}$ invariant. The only non-vanishing moduli fields $\varphi^a$ in the solution are required to describe a geodesic in $G_0/H_0$. The monodromy matrix is then chosen as follows:
\begin{equation}
    \mathfrak{M}\in G_0(\mathbb{Z})\,\subset\,{\rm O}(4,20;\,\mathbb{Z}) \,.
\end{equation}

\section{The Bosonic Sector of the  $\mathcal{N}=(2,2),\,D=6$ Theory} \label{bosonicsector}
In this section, we shall describe an instance of $U$-fold of the form ${\rm AdS}_2\times S^1\times S^3$, obtained applying the construction discussed in section \ref{GenConst} to the maximal six-dimensional model obtained from Type IIB supergravity upon compactification on a 4-torus and dualization of all forms are dualized to lower-order ones. 
As pointed out above, this model, in its classical limit,  features a scalar manifold of the form:
\begin{equation}
\mathcal{M}_{{\rm scal}}=\frac{{\rm SO}(5,5)}{{\rm SO}(5)\times {\rm SO}(5)}\,.\label{so55}
\end{equation}
Let us start reviewing the general mathematical description of the model.
We recall that the scalar fields originating from toroidal dimensional reduction to $D=6$ consist in the moduli of the internal metric $G_{ij}$, the axion-dilaton field $\rho=C_{(0)}+i\,e^{-\phi}$, the scalars $B^\upalpha_{ij}=(C_{ij},\,-B_{ij})$, $\upalpha=1,2$, originating from the ten-dimensional 2-forms $C_{(2)},\,B_{(2)}$ and the scalar $C_{ijkl}=c\,\epsilon_{ijkl}$ from the 4-form. As mentioned earlier, the five 2-form fields and their duals transform, under the global symmetry group ${\rm SO}(5,5)$, in its fundamental representation $\mathscr{R}_{(2)}={\bf 10}$.
They are $B^\upalpha_{\mu\nu}=(C_{\mu\nu},\,-B_{\mu\nu})$, their duals $\tilde{B}_{\upalpha\,\mu\nu}=(\tilde{C}_{\mu\nu},\,-\tilde{B}_{\mu\nu})$ and the six  $C_{ij\mu\nu}$. The identification of the $D=6$ scalar fields and 2-forms with the above Type IIB fields is effected by decomposing the adjoint and the fundamental representations of ${\rm SO}(5,5)$ with respect to the subgroup ${\rm SL}(4,\mathbb{R})\times {\rm SO}(1,1)\times {\rm SL}(2,\mathbb{R})_{{\rm IIB}}$, where ${\rm SL}(4,\mathbb{R})\times {\rm SO}(1,1)$ is the group of acting transitively on the metric moduli of the torus:
\begin{align}
{\rm Adj}({\rm SO}(5,5))&\rightarrow \,{\rm Adj}({\rm GL}(4,\mathbb{R}))+{\rm Adj}( {\rm SL}(2,\mathbb{R})_{{\rm IIB}})+{\bf (6,2)}_{+1}+{\bf (1,1)}_{+2}+\nonumber\\&+{\bf (6',2)}_{-1}+{\bf (1,1)}_{-2}\,,\nonumber\\
{\bf 10}&\rightarrow \,{\bf (1,2)}_{-1}+{\bf (6,1)}_{0}+{\bf (1,2)}_{+1}\,,
\end{align}
where the grading refers to the ${\rm SO}(1,1)$ factor. The scalars $B^\upalpha_{ij}$ parametrize the nilpotent generators in the ${\bf (6,2)}_{+1}$, $c$ the highest-grading generator ${\bf (1,1)}_{+2}$. As for the 2-forms, $B^\upalpha_{\mu\nu}$ and $\tilde{B}_{\upalpha\,\mu\nu}$ are defined by the ${\bf (1,2)}_{-1}$ and ${\bf (1,2)}_{+1}$ representations in the branching of the ${\bf 10}$, while $C_{ij\mu\nu}$ transform in the ${\bf (6,1)}_{0}$.\par
Besides these fields, the bosonic sector also consists of 16 vector fields which, however, will not play a role in our analysis.\par
Below we describe the subsector of the maximal D=6 theory describing the graviton, the scalar fields and the 2-forms. We shall collectively denote by $C_{(2)}^I$, $I=1,\dots, 5$ the tensor fields and by $\phi^s$ the 25 scalar fields.
The corresponding Lagrangian \eqref{lagrangian2d}) reads
\begin{align}
    \mathcal{L}_6= e_6 R_6 -\frac{e_6 }{12} \left(\mathcal{I}_{IJ} H^{I}_{\,\,\hat{\mu}\hat{\nu}\hat{\rho}}H^{J\,\hat{\mu}\hat{\nu}\hat{\rho}}+\mathcal{R}_{IJ} H^{I}_{\,\,\hat{\mu}\hat{\nu}\hat{\rho}} {}^\star H^{J\hat{\mu}\hat{\nu}\hat{\rho}}\right)- e_6 \frac{{\rm Tr}}{8}\left(\mathcal{M}^{-1}\partial \mathcal{M}\mathcal{M}^{-1}\partial \mathcal{M}\right), \end{align}
    where $\hat{\mu}=0,...,5$ and I,J=$1,...,n$. The matrices $\mathcal{I}(\phi)=(\mathcal{I}(\phi)_{IJ})=\mathcal{I}(\phi)^T>0$ and $\mathcal{R}(\phi)=(\mathcal{R}(\phi)_{IJ})=-\mathcal{R}(\phi)^T$ were introduced earlier in section \ref{GenConst}, together with the 
 symmetric, positive-definite ${\rm O}(5,5)$-matrix $(\mathcal{M})= \mathcal{M}_{MN}$, where $M,N=1,...,10$, see eq. \eqref{matrixM}.

\subsection{An Explicit Parametrization of the Scalar Manifold}
In this section, we review the explicit description of the scalar manifold $M_{{\rm scal.}}$ in terms of the so-called \emph{solvable} or \emph{Borel}  one \cite{Andrianopoli:1996zg,Cremmer:1997ct} parametrization, in which the Type IIB origin of the 25 scalar fields is manifest. A detailed account of this parametrization, for this model, is also given in \cite{Hull:2020byc}. Here we use, for the description of the Type IIB theory in $D=10$, the conventions defined in Appendix B of \cite{Guarino:2019oct}.\par
The basis of the $\mathscr{R}_{(2)}={\bf 10}$ is chosen as follows:\footnote{In this representation the coefficient ${\tt k}$ in \eqref{MdMg} is equal to 1.}
\begin{equation}
V^M=(V^\alpha,\,V_{ij},\,V_\alpha)\,,\label{10b}
\end{equation}
where the components $V^\alpha,\,V_{ij},\,V_\alpha$ transform in the ${\bf (1,2)}_{-1}$, ${\bf (6,1)}_{0}$ and ${\bf (1,2)}_{+1}$, respectively. In this basis, the ${\rm SO}(5,5)$ invariant matrix in the fundamental reads:
\begin{equation}
\Omega_{MN}\equiv \left(\begin{matrix}{\bf 0} & {\bf 0} & \delta^\alpha_\beta\cr {\bf 0} &\epsilon_{ij\,kl} &{\bf 0}\cr
\delta^\alpha_\beta&{\bf 0} & {\bf 0} \end{matrix}\right)\,.
\end{equation}
In the solvable parametrization of the scalar manifold the scalar fields are defined as parameters of the Borel subalgebra of $\mathfrak{so}(5,5)$.\par
 We start by discussing the axion-dilaton system and then describe the rest of the scalar fields.\par
\paragraph{Axion-dilaton system.} The dilaton $\phi$ and the axion field $C_{(0)}$, in the ten-dimensional theory, span the coset ${\rm SL}(2,\mathbb{R})_{{\rm IIB}}/{\rm SO}(2)$. The corresponding coset representative is chosen of the following form:
\begin{equation}
L_{(2)}=(L_\upalpha{}^{\upbeta})=\left(
\begin{array}{cc}
 e^{\phi /2} & 0 \\
 C_{(0)}\,  e^{\phi /2} & e^{-\phi /2} \\
\end{array}
\right)\,.\label{L2}
\end{equation}
This allows us to define the matrix
\begin{equation}
\mathbf{m}_{\upalpha\upbeta}\equiv (L_{(2)}L_{(2)}^T)_{\upalpha\upbeta}=\frac{1}{{\rm Im}(\rho)}\left(\begin{matrix}1 & {\rm Re}(\rho)\cr {\rm Re}(\rho) & |\rho|^2 \end{matrix}\right)\,.
\end{equation}
\paragraph{The ${\rm SO}(5,5)/{\rm SO}(5)\times {\rm SO}(5)$ manifold.}
As mentioned above, the scalars $B^\upalpha_{ij}$ and $c$ parametrize the generators $t_\upalpha^{ij},\,t$ of the Borel subalgebra of $\mathfrak{so}(5,5)$ in the representations ${\bf (6,2)}_{+1}$ and ${\bf (1,1)}_{+2}$ of ${\rm GL}(4,\,\mathbb{R})$, respectively.
In the basis \eqref{10b}, these generators read:
\begin{align}
\mathbb{B}&=\frac{1}{2}\,B^\upalpha_{ij}\,t_\upalpha^{ij}=\left(\begin{matrix}{\bf 0} & {\bf 0} & {\bf 0}\cr B_{\upalpha ij} &{\bf 0} &{\bf 0}\cr
{\bf 0} &-B_{\upalpha}{}^{ij} & {\bf 0} \end{matrix}\right)\,,\,\,\,\,\,\mathbb{C}=c\,t=\left(\begin{matrix}{\bf 0} & {\bf 0} & {\bf 0}\cr {\bf 0}&{\bf 0} &{\bf 0}\cr
c\,\epsilon_{\upalpha\upbeta} & {\bf 0} & {\bf 0} \end{matrix}\right)\,,
\end{align}
where
\begin{equation}
B_{\upalpha\,ij}\equiv\,\epsilon_{\upalpha\upbeta}\,B^\beta_{ij}=(B_{ij},\,C_{ij})\,,\,\,\,\,B_{\upalpha}{}^{ij} \equiv \frac{1}{2}\,\epsilon^{ijkl}\,B_{\upalpha\,kl}\,.
\end{equation}
Next we write the coset representatives $\mathbb{L}_{(2)},\,\mathbb{L}_{(4)}$ of ${\rm SL}(2,\mathbb{R})_{{\rm IIB}}/{\rm SO}(2)$ and ${\rm GL}(4,\mathbb{R})/{\rm SO}(4)$ in the same representation:
\begin{align}
\mathbb{L}_{(2)}&=\left(\begin{matrix}L_{(2)}^{-T} & {\bf 0} & {\bf 0}\cr {\bf 0}&{\bf 1} &{\bf 0}\cr
{\bf 0} & {\bf 0} & L_{(2)} \end{matrix}\right)\,,\nonumber\\
\mathbb{L}_{(4)}&=\left(\begin{matrix}G^{-\frac{1}{4}}\,{\bf 1} &{\bf 0}& {\bf 0}\cr {\bf 0}&2\, G^{-\frac{1}{4}}\,E_{[i}{}^k E_{j]}{}^k &{\bf 0}\cr
{\bf 0} & {\bf 0} &G^{\frac{1}{4}}\,{\bf 1} \end{matrix}\right)\,,
\end{align}
where $G\equiv {\rm det}(G_{ij})$, the metric $G_{ij}$ being in the $D=10$ Einstein frame and $E_i{}^k$ being the vielbein matrix: $G_{ij}=E_i{}^kE_j{}^k$ (summation over $k$ is understood).
The $\frac{{\rm SO}(5,5)}{{\rm SO}(5)\times {\rm SO}(5)}$ coset representative is then written as follows:
\begin{equation}
\mathbb{L}^M{}_N=e^{\mathbb{C}} \cdot e^{\mathbb{B}} \cdot \mathbb{L}_{(4)}  \cdot\mathbb{L}_{(2)}\,.\label{cosetr}
\end{equation}
We can now compute the components of $\mathcal{M}=\mathbb{L}\, \mathbb{L}^T$:
\begin{align}
\mathcal{M}^{\upalpha\upbeta}&=G^{-\frac{1}{2}}\,\mathbf{m}^{\upalpha\upbeta}\,\,,\nonumber\\
\mathcal{M}_{ij}{}^{\upbeta}&=G^{-\frac{1}{2}}\,\mathbf{m}^{\upalpha\upbeta}\,B_{\upalpha\,ij}\,\,,\nonumber\\
\mathcal{M}_\upalpha{}^\upbeta&=G^{-\frac{1}{2}}\,c_{\upalpha\upsigma}\,\mathbf{m}^{\upsigma\upbeta}\,\,,\nonumber\\
\mathcal{M}_{ij,\,kl}&=G^{-\frac{1}{2}}\,\left(2\,G_{i[k} G_{l]j}+\mathbf{m}^{\upalpha\upbeta}\,B_{\upalpha\,ij}\,B_{\upbeta,kl}\right)\,,\nonumber\\
\mathcal{M}_{\upalpha\,ij}&=G^{-\frac{1}{2}}\,\left( c_{\upalpha\upbeta}\,\mathbf{m}^{\upbeta\upsigma}\,B_{\upsigma\,ij}-B_\upalpha{}^{kl}\,G_{ki} G_{lj}\right)\,,\nonumber\\
\mathcal{M}_{\upalpha\upbeta}&=G^{-\frac{1}{2}}\,\left(c_{\upalpha\upsigma}c_{\upbeta\upgamma}\,\mathbf{m}^{\upsigma\upgamma}+\frac{1}{2}\,B_{\upalpha}{}^{ij}B_{\upbeta}{}^{kl}\,G_{ik}G_{jl}+G\,\mathbf{m}_{\upalpha\upbeta}\right)\,,\end{align}
where:
\begin{equation}
c_{\upalpha\upbeta}\equiv c\,\epsilon_{\upalpha\upbeta}-\frac{1}{4}\,B_{\upalpha\,ij}B_\upbeta{}^{ij}\,.
\end{equation}
We refer to the Appendix \ref{AppendixA} for more details about the solvable parametrization of the scalar manifold as well as the description of the tensor fields in terms of the $\mathfrak{so}(5,5)$ weights of the ${\bf 10}$.

\subsection{${\rm AdS}_3\times { S}^3$ Solution of Type IIB Theory on $T^4$}\label{ReviewAdS3S3}
In the appendix \ref{Stringsolutionsapp} we present a general discussion of the static black-string solutions in the presence of non-trivial moduli. A known instance is the D1-D5 system in which the D1 and D5-branes have only two common Neumann directions, defining the worldsheet of the black string in $D=6$, and the D5-branes wrap the whole $T^4$. 
This solution is of particular importance in string theory since it provided the setting for the first, successful black hole entropy calculation through string microstate counting \cite{Strominger:1996sh}.\par
The charge vector reads 
\begin{equation}
    \Gamma^M=(d_5,0,0,0,0,0,0,0,d_1,0)\,,\label{d1d5G}
\end{equation}
where $d_5,\,d_1$ are taken to be both positive and denote the D5 and the D1-brane charges, respectively. Through a (charge-dependent) change of basis, see Appendix \ref{CriticalPointsPot}, the above vector can be brought to the form:
\begin{equation}
    \Gamma^{\prime M}=(\sqrt{2 |d_1 d_5|},0,0,0,0,0,0,0,0,0)\,.
\end{equation}
In this new basis, the invariant pseudo-orthogonal metric reads:
\begin{equation}
    \Omega'={\rm diag}(+1,+1,+1,+1,+1,-1,-1,-1,-1,-1)\,.
\end{equation}
The above form of $\Gamma'$ makes its little group $G_\ell={\rm SO}(4,5)$ manifest: it is the subgroup of $G$ acting non-trivially only on the entries $2,3,\dots, 10$. The matrix representation of $G_\ell$ in the original basis \eqref{d1d5G} is $d_1,\,d_5$-dependent. As explained in sections \ref{GenConst} and \ref{examples}, the coset space $\mathbb{K}$ of $M_{{\rm scal.}}$ splits into $\mathbb{K}_\ell$ and $\hat{\mathbb{K}}$ supporting the representations ${\bf (4,5)}$ and ${\bf (1,5)}$ of $H_\ell={\rm SO}(4)\times {\rm SO}(5)$. The non-moduli fields $\phi^k$ parametrize the five-dimensional $\hat{\mathbb{K}}$ and can be brought to a normal form $\phi^k=(g,0,0,0,0)$, through the action of ${\rm SO}(5)$, in which the only non-vanishing scalar $g$ parametrizes the Cartan generator:
\begin{equation}
    \mathfrak{h}={\rm diag}(1,0,0,0,0,0,0,0,-1,0)\,.\label{frakh}
\end{equation}
The scalar field $g$ is identified as follows: $e^g=e^{\phi}\,G^{\frac{1}{2}}$, where $G\equiv {\rm det}(G_{ij})$, see Appendix \ref{AppendixA}. \par
The effective potential $V$ for the D1-D5
 system, and its extremization, is discussed in Appendix \ref{CriticalPointsPot}.
 The maximal subgroup of ${\rm SO}(5,5)$ commuting with $\mathfrak{h}$ is ${\rm SO}(4,4)$, which represents the maximal choice of $G_0\subset G_\ell$. The maximal moduli space $M_0$ is:
 $$M_0=\frac{{\rm SO}(4,4)}{{\rm SO}(4)\times {\rm SO}(4)}\,,$$
 parametrized by $C_{ij}$ and $\tilde{G}_{ij}=e^{-\frac{\phi}{2}}\,G_{ij}$, see Appendix \ref{AppendixA}.
 Had we chosen the system of F1 and NS5 branes
dual to the D1-D5 one, the maximal moduli space $M_0$ would be spanned by $B_{ij},\,G^{(s)}_{ij}\equiv e^{\frac{\phi}{2}}\,G_{ij}$, where $G^{(s)}_{ij}$ is the torus metric in the string frame. The ${\rm O}(4,4;\,\mathbb{Z})$ global symmetry group would, in this case, encode the T-dualities along directions of $T^4$, and transformations in ${\rm O}(4,4;\,\mathbb{Z})/{\rm SO}(4,4;\,\mathbb{Z})$ would map the Type IIB and Type IIA descriptions of the same F1-NS5 solution.

Coming back to the D1-D5 system, when we only switch on the moduli fields $C_{ij}$,  $\tilde{G}_{ij}$ and the field $g$, 
the matrix $\mathcal{M}$ is the product of two commuting symmetric matrices:
\begin{equation}
\mathcal{M}(\varphi^a,\,g)=\mathcal{M}_0(\varphi^a)\,\hat{\mathcal{M}}(g)\,,
\end{equation}
where
\begin{align}
  \hat{\mathcal{M}}(g)&=  \text{diag}(e^g,1,1,1,1,1,1,1,1,e^{-g},1)\,,
\end{align}
and $\Gamma^T\,\mathcal{M}(\varphi^a,g)\Gamma=\Gamma^T\,\hat{\mathcal{M}}(g)\Gamma$, so that 
the effective potential reads:
\begin{equation}
    V=\frac{1}{2}\left(d_1^2\, e^{-g}+d_5^2\, e^{g}\right)\,.\label{Vg}
\end{equation}
By further fixing the moduli fields $C_{ij}=0,\,\tilde{G}_{ij}=\delta_{ij}$, we have $g=2\phi$ and the solution takes the known form (see appendix \ref{D1-D5})
    \begin{align}
    ds^2 =& (Z_1 Z_5)^{-\frac{1}{2}} (-dt^2+dx^2)+ (Z_1 Z_5)^{\frac{1}{2}} (dx^i dx^i)\,, \\
    e^g&=e^{2 \phi}=\log\left(\frac{Z_1}{Z_5}\right)\,,\nonumber\\
    dx^i dx^i =& dr^2 + r^2 d\Omega^2_3, \qquad Z_1=1+ \frac{d_1}{2r^2},\qquad Z_5= 1+\frac{d_5}{2r^2}.
\end{align}
To avoid the factor $1/2$ in the harmonic functions, it can be convenient to define rescaled charges: $d_1=2 Q_1,\,d_5=2 Q_5$. The charges $d_1$ and $d_5$ are defined in eq.(\ref{ChargesDef}).  At the horizon the $Z$ functions become
\begin{align}
   Z_1= \frac{d_1}{2r^2}=\frac{Q_1}{r^2},\qquad Z_5= \frac{d_5}{2r^2}=\frac{Q_5}{r^2}.
\end{align}
This background is AdS$_3 \times S^3$, which is the throat of the D1-D5 system. The non-moduli scalars are attracted towards configurations that extremizes the potential.  We give a general discussion in appendix \ref{CriticalPointsPot}. In this particular case, in which we set $C_{ij}=0,\,\tilde{G}_{ij}=\delta_{ij}$, the scalar field $g=2\phi$ is fixed at the horizon, through the attractor mechanism, to the value
\begin{align}
  g= 2\phi= \log\left(\frac{d_1}{d_5}\right)= \log\left(\frac{Q_1}{Q_5}\right)\,.
\end{align}
At the minimum the effective potential has the value
\begin{equation}
    V_*=d_1 d_5\,.\label{ell}
\end{equation}
The near-horizon metric is conveniently written in the following form:
\begin{gather}
    ds^2= \frac{r^2}{L^2}(-dt^2+dx^2)+ \frac{L^2}{r^2} dr^2+ L^2 \left[d\psi^2 + \sin^2\psi\left(d\theta^2+\sin^2\theta d\omega^2\right) \right],
\end{gather}
where $L^2=\frac{\sqrt{d_1 d_5}}{2}$, as obtained in eqs.\eqref{adsdsdcase}, \eqref{ell}.

\subsection{${\rm AdS}_2\times S^1\times { S}^3$ U-Folds} \label{U-Foldsection} 
In this section, we apply the general construction of section \ref{GenConst} to the construction of $U-$fold solutions with geometry  AdS$_2\times S^1 \times S^3$ in
Type IIB theory on $T^4$. As discussed in section \ref{GenConst}, we only switch on the field $e^g=e^{\phi}\,G^{\frac{1}{2}}$ and the moduli fields $\varphi^a$ spanning either the manifold ${\rm SO}(4,4)/[{\rm SO}(4)\times {\rm SO}(4)]$, parametrized by $\tilde{G}_{ij}=e^{-\frac{\phi}{2}}G_{ij},\,C_{ij}$, or a symmetric submanifold $M_0=G_0/H_0$ of it.
We start with a metric of the form
\begin{gather}
    ds^2= -k_1\frac{r^2}{L^2}dt^2+ k_1\frac{L^2}{r^2} dr^2+k_2 d\eta^2+L^2 \left[d\psi^2 + \sin^2\psi\left(d\theta^2+\sin^2\theta d\omega^2\right) \right],
\end{gather}
where $k_1,\,k_2$ are positive constants and we compactify the $\eta$ direction: $\eta\in [0,T]$.
Just as in the D1-D5 solution, we turn on the charges $d_1,\,d_5$. 
The ansatz for the three form field strength is again 
\begin{align}
    \mathbb{H}=- \gamma \Omega \mathcal{M} \Gamma dt \wedge d\eta \wedge dr+ \zeta(\theta,\psi) \Gamma d\psi\wedge d\theta \wedge d\omega, \label{U-Fold-Hstrenght}
\end{align}
which is duality-covariant. On the 3-form field
we need to impose the twisted self-duality condition, the Maxwell equations, and the Bianchi equations, which read respectively
\begin{align}
    \star \mathbb{H}= - \Omega\, \mathcal{M} \mathbb{H}, \quad
    d\mathbb{H}=0\,. 
\end{align}
These are all solved if we pick
\begin{align}
    \gamma=\frac{k_1 \sqrt{k_2}}{L^3}, \quad \zeta(\theta,\psi)= \sin \theta \sin^2 \psi.
\end{align}
The moduli $\varphi^a$ in  $M_0$ are switched on along a geodesic arc:
\begin{equation}
    \varphi^a=\varphi^a(\eta)\,,
\end{equation}
with "velocity" $\kappa$ and affine parameter $\eta$, so that
\begin{align}
    \frac{1}{2}\mathcal{G}_{ab} \partial_\eta \varphi^a \partial_\eta \varphi^b= \kappa^2.\label{unitmeasure}
\end{align}
Thanks to the ansatz for the field strength $\mathbb{H}$ (\ref{U-Fold-Hstrenght}), the equation of motion for the scalars has the same form as eq.(\ref{ScalarEqGen}), where $V(\phi,\Gamma)=V(g,\Gamma)$ is given in eq. \eqref{Vg}.
The scalar equation of motion is solved if we extremize the potential as
\begin{align}
  g= \log\left(e^{\phi}\,G^{\frac{1}{2}}\right)= \log\left(\frac{d_1}{d_5}\right)={\rm const.},
\end{align}
since we forced the moduli fields $\varphi^a$ to evolve in $\eta$ along a geodesic. \\
The last equation to check is the trace-reversed Einstein equation. The computation of the energy-momentum part which involves the three form $\mathbb{H}$ gives
\begin{align}
   T^{(H)}_{\hat{\mu}\hat{\nu}}=\frac{d_1 d_5}{2 L^4}\,{\rm diag}\left(\frac{k_1 r^2}{L^4},-\frac{k_2}{L^2},-\frac{k_1}{r^2},1,\sin ^2(\psi ),\sin ^2(\theta ) \sin ^2(\psi )\right)\,,
\end{align}
where the second element on the diagonal is the $(\eta,\eta)$ component of the tensor and has to be canceled, in the Einstein equation, by the only non-vanishing component of the scalar energy-momentum tensor, which equals the left-hand side of eq.\eqref{unitmeasure}, namely $\kappa^2$. The Einstein equation then completely fixes the coefficients as follows
\begin{align}
    k_1=\frac{1}{2}, \quad k_2= \frac{\kappa^2\,\sqrt{d_1 d_5}}{4},\quad L^2=\frac{\sqrt{d_1 d_5}}{2}.
\end{align}
To give an instance of a geodesic curve described by string 0-modes, we can restrict the moduli fields $\varphi^a$ to the submanifold
\begin{equation}
    M_0=\frac{G_0}{H_0}=\left(\frac{{\rm SL}(2,\mathbb{R})}{{\rm SO}(2)}\right)^2\,,
\end{equation}
which is obtained by restricting the moduli $\tilde{G}_{ij}=e^{-\frac{\phi}{2}}\,G_{ij},\,C_{ij}$ to the following non-vanishing fields:
\begin{equation}
    e^{-\phi_{12}}\equiv \tilde{G}_{11}=  \tilde{G}_{22}\,,\,\,e^{-\phi_{34}}\equiv \tilde{G}_{33}=  \tilde{G}_{44}\,,\,\,\,C_{12},\,C_{34}\,.
\end{equation}
The metric of the scalar manifold, restricted to $g$, $\phi_{12},\,\phi_{34},\,C_{12},\,C_{34}$ reads:
\begin{equation}
    d\tilde{s}^2=\frac{dg^2}{2}+d\phi_{12}^2+d\phi_{34}^2+e^{2\phi_{12}}\,dC_{12}^2+e^{2\phi_{34}}\,dC_{34}^2\,. \label{subsectortargetspace}
\end{equation}
The geodesic can be chosen as the product of two geodesics spanned by $\phi_{12},\,C_{12}$ and by $\phi_{34},\,C_{34}$, respectively, unfolding in the two factors of $M_0$.
The monodromy matrix $\mathfrak{M}$ will be the product $\mathfrak{M}=\mathfrak{M}_1\,\mathfrak{M}_2$, of two elements $\mathfrak{M}_1,\,\mathfrak{M}_2$ in the two factors ${\rm SL}(2,\mathbb{Z})$ in $G_0(\mathbb{Z})={\rm SL}(2,\mathbb{Z})^2$.
Choosing, for instance:
$$\mathfrak{M}_1=J_{{\tt n}_1}\,,\,\,\,\,\mathfrak{M}_2=J_{{\tt n}_2}\,,$$
the geodesic reads:
 \begin{align}
    e^{\phi_{12}(\eta)}&=\frac{\sqrt{{\tt n}_1 ^2+4} \cosh \left(\frac{\eta  \cosh ^{-1}\left(\frac{1}{2} \left({\tt n}_1 ^2+2\right)\right)}{T}\right)+{\tt n}_1  \sinh
   \left(\frac{\eta  \cosh ^{-1}\left(\frac{1}{2} \left({\tt n}_1 ^2+2\right)\right)}{T}\right)}{\sqrt{{\tt n}_1 ^2+4}}\,,\nonumber\\
   C_{12}(\eta)&=-\frac{2 \sinh \left(\frac{\eta  \cosh ^{-1}\left(\frac{1}{2} \left({\tt n}_1 ^2+2\right)\right)}{T}\right)}{\sqrt{{\tt n}_1 ^2+4} \cosh
   \left(\frac{\eta  \cosh ^{-1}\left(\frac{1}{2} \left({\tt n}_1 ^2+2\right)\right)}{T}\right)+{\tt n}_1  \sinh \left(\frac{\eta  \cosh
   ^{-1}\left(\frac{1}{2} \left({\tt n}_1 ^2+2\right)\right)}{T}\right)}\,,\nonumber\\
       e^{\phi_{34}(\eta)}&=\frac{\sqrt{{\tt n}_2 ^2+4} \cosh \left(\frac{\eta  \cosh ^{-1}\left(\frac{1}{2} \left({\tt n}_2 ^2+2\right)\right)}{T}\right)+{\tt n}_2  \sinh
   \left(\frac{\eta  \cosh ^{-1}\left(\frac{1}{2} \left({\tt n}_2 ^2+2\right)\right)}{T}\right)}{\sqrt{{\tt n}_2 ^2+4}}\,,\nonumber\\
   C_{34}(\eta)&=-\frac{2 \sinh \left(\frac{\eta  \cosh ^{-1}\left(\frac{1}{2} \left({\tt n}_2 ^2+2\right)\right)}{T}\right)}{\sqrt{{\tt n}_2 ^2+4} \cosh
   \left(\frac{\eta  \cosh ^{-1}\left(\frac{1}{2} \left({\tt n}_2 ^2+2\right)\right)}{T}\right)+{\tt n}_2  \sinh \left(\frac{\eta  \cosh
   ^{-1}\left(\frac{1}{2} \left({\tt n}_2 ^2+2\right)\right)}{T}\right)}\,,\nonumber
\end{align}
which satisfy the equations explicitly presented in appendix \ref{Stringsolutionsapp}, after the appropriate coordinate transformations to bring the target metric (\ref{subsectortargetspace}) in the form (\ref{metricSL2}). It is straightforward to verify that, as $\eta\rightarrow \eta+T$ the complexified fields $\rho_{12}\equiv C_{12}+i\,e^{-\phi_{12}},\,\rho_{34}\equiv C_{34}+i\,e^{-\phi_{34}}$ transform under the action of the monodromies $J_{{\tt n}_1}$, $J_{{\tt n}_2}$, respectively:
$$\rho_{12}\rightarrow -\frac{1}{\rho_{12}+{\tt n}_1}\,\,,\,\,\,\rho_{34}\rightarrow -\frac{1}{\rho_{34}+{\tt n}_2}\,.$$
In this case, we have:
\begin{align}
    \kappa^2=\frac{1}{2T^2}\left(\cosh^{-1}\left(\frac{{\tt n}_1 ^2}{2}+1\right)+\cosh^{-1}\left(\frac{{\tt n}_2 ^2}{2}+1\right)\right)\,.
\end{align}
Along the same lines, we can construct the ${\rm AdS}_2\times S^1\times S^3$ T-fold starting from the near horizon geometry of an F1-NS5 system with charges 
$$\Gamma^M=(0,n_5,0,0,0,0,0,0,0,f_1)\,.$$
This solution is obtained by applying an S-duality $\mathcal{S}\in {\rm SL}(2,\mathbb{Z})_{{\rm IIB}}$ to the D1-D5 U-fold. As pointed out earlier, the maximal choice ${\rm SO}(4,4)/{\rm SO}(4)\times {\rm SO}(4)$ for the moduli space $M_0$ is spanned by $B_{ij},\,G^{(s)}_{ij}\equiv e^{\frac{\phi}{2}}\,G_{ij}$, where $G^{(s)}_{ij}$ is the torus metric in the string frame, while 
\begin{equation}\hat{\mathcal{M}}={\rm diag}(1,e^g,1,1,1,1,1,1,1,e^{-g})\,,\label{NSg}\end{equation}
with $e^g=e^{-2\phi_6}\equiv e^{-\phi}\,G^{\frac{1}{2}}$, see Appendix \ref{AppendixA}.
Since now the monodromy matrix is an element of ${\rm O}(4,4; \mathbb{Z})$, which is now the T-duality group associated with the internal 4-torus, the resulting U-fold is a T-fold.\par We can choose a monodromy matrix $\mathfrak{M}\in {\rm O}(4,4; \mathbb{Z})/{\rm SO}(4,4; \mathbb{Z})$, which is not connected continuously to the identity. This can be achieved by describing $S^1$ through two open patches: one consisting of $S^1$ minus the point $\eta=0$, the other obtained by subtracting to $S^1$ the point $\eta=T/2$. Within each patch the evolution of the scalar fields is geodesic. However, in the interval $(0,T/2)$ the moduli fields are described in both patches by\footnote{Recall that $\mathbb{L}_0$ is in the identity sector of the isometry group.}  $\mathbb{L}_0(\varphi^a(\eta))$ while in $(T/2,T)$ they are described in the first patch by $\mathbb{L}_0(\varphi^a(\eta))$ and in the  second one by $\mathscr{O}\cdot \mathbb{L}_0(\varphi^a(\eta))$, where $\mathscr{O}\in {\rm O}(4,4; \mathbb{Z})/{\rm SO}(4,4; \mathbb{Z})$, $\mathscr{O}^2={\bf 1}$. The twist matrix has the following property: $\mathcal{A}(0)=\mathbb{L}_0(\varphi^a(0))$, $\mathcal{A}(T)=\mathscr{O}\cdot\mathbb{L}_0(\varphi^a(T))$, so that $\mathfrak{M}$, given by eq. \eqref{MonodromyM}, is in ${\rm O}(4,4; \mathbb{Z})/{\rm SO}(4,4; \mathbb{Z})$.  
This T-fold solution with NS-NS charges is therefore described in one patch within the Type IIB theory and in the other one within the Type IIA one, the transition function being a pseudo-orthogonal matrix with a negative determinant.\par
In the next subsection, we review the action of a duality transformation on the general, duality covariant, ansatz discussed in section \ref{GenConst}.

\subsubsection{Duality}
\label{duality}
Let us discuss here the action of the duality symmetry on a generic solution.
Consider a duality transformation 
\begin{equation}
    \mathcal{O}=(\mathcal{O}^M{}_N)\in {\rm SO}(5,5)\,\,\Leftrightarrow\,\,\mathcal{O}\,\Omega\,
    \mathcal{O}^T=\Omega\,.
\end{equation}
Its action on the scalar fields is described by the transformation property of the matrix $\mathcal{M}(\phi)$:
\begin{equation}
    \phi\equiv (\phi^s)\,\stackrel{\mathcal{O}}{\longrightarrow}\,\,\phi^\prime\equiv (\phi^{\prime s})\,\,\Leftrightarrow\,\,\,\,\mathcal{O}^{-T}\mathcal{M}(\phi)\mathcal{O}^{-1}=\mathcal{M}(\phi')\,,\label{phip}
\end{equation}
while the corresponding action on $\mathbb{H}\equiv (\mathbb{H}^M)$ reads:
\begin{align}
    \mathbb{H}'= \mathcal{O}\,\mathbb{H}= -\Omega \mathcal{O}^{-T} \mathcal{M}(\phi)\mathcal{O}^{-1} \mathcal{O} \Gamma \boldsymbol{\epsilon}_{(3)}+\mathcal{O} \Gamma \Tilde{\boldsymbol{\epsilon}}_{(3)}=-\Omega \mathcal{M}(\phi') \Gamma' \boldsymbol{\epsilon}_{(3)}+ \Theta \Gamma' \Tilde{\boldsymbol{\epsilon}}_{(3)}\,.
\end{align}
A particular instance of duality symmetry, which we shall use in the sequel, is $S$-duality, implemented, in the basis we are using, by the following matrix
\begin{align}
    \Theta&=\left(\begin{matrix}{\theta^{\alpha}_{\,\,\beta} } & {\bf 0} & {\bf 0}\cr {\bf 0} & 2 \delta_{ij}^{\,\,kl} &{\bf 0}\cr
{\bf 0} &{\bf 0} & {\tilde{\theta}_{\alpha}^{\,\,\beta}} \end{matrix}\right),
\end{align}
where
\begin{align}
    \theta^{\alpha}_{\,\,\beta} =\left(\begin{matrix}
    0 & 1 \cr -1 &0
    \end{matrix}\right), \quad \tilde{\theta}_{\alpha}^{\,\,\beta}=\left(\begin{matrix}
    0 & 1 \cr -1 &0
    \end{matrix}\right),
\end{align}
such that $\tilde{\theta}_{\alpha}^{\,\,\beta}=(\theta^{-1})^{\beta}_{\,\,\alpha}$. Besides being pseudo-orthogonal, $\Theta$ is orthogonal as well: $\Theta\,\Theta^T={\bf 1}.$
From eq. \eqref{phip} we deduce the following transformation rule for the scalar fields:
\begin{align}
    B'_{\alpha}&=\tilde{\theta}_{\alpha}^{\,\,\beta} B_{\beta}= \left(\begin{matrix}
    0 &1 \cr -1 & 0
    \end{matrix}\right) \left(\begin{matrix}
    B_1 \cr B_2
    \end{matrix}\right)= \left(\begin{matrix}
    B_2 \cr -B_1
    \end{matrix}\right), \label{Sdual1} \\
    c'\epsilon_{\alpha\sigma}&=c (\theta^{-1})^{\beta}_{\,\,\alpha} \epsilon_{\beta \gamma} \tilde{\theta}_{\sigma}^{\,\,\gamma}=c \epsilon_{\alpha\sigma} \label{Sdual2}, \\
    \rho'&=-\frac{1}{\rho} \label{Sdual3},
\end{align}
all other fields being invariant.

\section{Final remarks and discussion of the solution}
In this work, we discuss the general construction of a U-fold solution with ${\rm AdS}_{d-1}\times S^1 \times S^d$ geometry and a monodromy along $S^1$ in the duality group, within theories admitting a ${\rm AdS}_d \times S^d$ background with moduli fields. More in general we build U-fold backgrounds of the form
\begin{align}
    {\rm AdS}_{d-k}\times \underbrace{S^1\times \dots \times S^1}_k\times S^d\,,
\end{align}
with $d-k>1$.   \par
This was achieved by introducing an appropriate spatial dependence of the scalar moduli along a compact direction which, in the original ${\rm AdS}_{d}\times S^d$ solution, is in the boundary of ${\rm AdS}_{d}$.
The effect of the back-reaction of the evolving moduli on the geometry is to split ${\rm AdS}_{d}$ into ${\rm AdS}_{d-1}\times S^1$. The geodesic arc connects two distinct points in the classical moduli space. Consequently, the consistency of this background in superstring theory requires the initial and ending points to be identified in the string moduli space. \par
Since the dependence of the fields, in the U-fold solution, on the $S^1$-coordinate $\eta$ is factorized in the action of a twist matrix $\mathcal{A}(\eta)\in G_0$ on the fields, it is natural to expect this background to be a solution of the $D=2d$-dimensional model compactified along $S^1$ \'a la Cremmer-Scherk-Schwarz \cite{Cremmer:1979uq} to $(2d-1)$ dimensions, with a twist matrix $\mathcal{A}(\eta)$ in the global symmetry group $G_0$. Indeed the only fields in the solution transforming non-tivially under $G_0$ are the moduli fields $\varphi^a$.
We extensively discussed, as a concrete application of our general construction, Type IIB superstring theory compactified on $T^4$ to six dimensions, and utilized the moduli fields present in the ${\rm AdS}_3\times { S}^3\times T^4$ solution to create a U-fold with geometry ${\rm AdS}_2\times S^1\times {S}^3\times T^4$. This construction involves introducing a non-trivial geodesic dependence of the moduli fields in ${\rm SO}(4,4)/{\rm SO}(4)\times {\rm SO}(4)$, or in a symmetric submanifold $M_0$ thereof, on a compact direction $S^1$ along the boundary of ${\rm AdS}_3$. \par
 Cremmer-Scherk-Schwarz reduction of maximal $\mathcal{N}=(2,2)$ six-dimensional supergravity to five dimensions admit Minkowski five-dimensional vacua only if the twist matrix is compact, that is in ${\rm Spin}(5)\times {\rm Spin}(5)$. In this case the reduction was originally  studied in \cite{deWit:2002vt}, \cite{Andrianopoli:2004xu} and, more recently, in \cite{Hull:2020byc,Gkountoumis:2023fym} in relation to the study of black hole solutions.
 Our setting, however, is substantially different since it relies on the moduli fields describing a geodesic in $M_0=G_0/H_0$, which requires the twist matrix $\mathcal{A}(\eta)\in G_0/H_0$ to be intrinsically non-compact. This is necessary for the evolving moduli to back-react on the geometry through a non-vanishing component of the energy-momentum tensor: $T^{(S)}_{\eta\eta}\neq 0$.\par
 The general mechanism analyzed here in detail, underlying the construction of ${\rm AdS}_{d-1}\times S^1 \times S^d$ U-folds, seems to be at work in the more sophisticated class of J-fold solutions with geometry ${\rm AdS}_4\times S^1\times \tilde{S}^5$, which have recently attracted considerable interest \cite{Inverso:2016eet,Assel:2018vtq,Guarino:2019oct,Guarino:2020gfe,Giambrone:2021zvp,Guarino:2021kyp,Arav:2021gra,Bobev:2021yya,Bobev:2020fon,Cesaro:2021tna,Guarino:2021hrc,Bobev:2021rtg,Guarino:2022tlw}. All these solutions correspond to ${\rm AdS}_4$ vacua of maximal four-dimensional supergravity with gauge group $\mathcal{G}=[{\rm SO}(6)\times {\rm SO}(1,1)]\ltimes T^{12}$. Aside from the non-supersymmetric J-fold with ${\rm SO}(6)$ symmetry mentioned earlier and found in \cite{Guarino:2019oct}, which fits the general class discussed in section \ref{GenConst}, in all other solutions the the axion-dilaton system features a dependence on the coordinates $\upxi^i$ of the deformed 5-sphere $\tilde{S}^5$ and on $\eta$. The former is sourced by 3-form field strengths while the latter describes a geodesic arc in the moduli space, as noticed in \cite{Arav:2021gra}. In the present work, we have highlighted this common feature of all these solutions and generalized it to the case $d\neq 5$, showing that the geodesic dependence of the moduli fields on the $\eta$ coordinate of an $S^1$ at the boundary of ${\rm AdS}_d$ is sufficient to split this space into ${\rm AdS}_{d-1}\times S^1$, where the ratio of the warp factors multiplying the metric  ${\rm AdS}_{d-1}$ and $d\eta^2$ is a suitable constant.\par
 An important issue to address is the supersymmetry and stability of the ${\rm AdS}_2\times S^1\times S^3$ solution discussed in this work. Secondly, it would be of considerable interest to construct variants of these solutions, with geometry ${\rm AdS}_2\times S^1\times \tilde{S}^3$, where $\tilde{S}^3$ is a deformed 3-sphere, in which the moduli fields $\varphi^a$, aside from the geodesic dependence on $\eta$, also feature a dependence on the coordinates of $\tilde{S}^3$ sourced by 2-form field-strengths. These new solutions would be the counterpart in $D=6$ of the $\mathcal{N}=1,2$ and $4$ J-folds in $D=10$. In analogy with the latter solutions, we may argue that these ${\rm AdS}_2\times S^1\times \tilde{S}^3$ U-folds should be dual to the IR limit of a $1+1$ SCFT compactified along one spatial direction, with an interface localized along this direction and an ${\rm O}(4,4;\,\mathbb{Z})$ monodromy acting on the exactly marginal operators dual to $\varphi^a$, when crossing the interface.
Although we do not expect the solutions ${\rm AdS}_{2}\times S^1\times S^3$ to be supersymmetric, we expect their variants ${\rm AdS}_{2}\times S^1\times \tilde{S}^3$ to preserve some supersymmetry, in analogy with the $D=10$ J-fold counterparts.
It would also be interesting to relate generalizations of our backgrounds to a compactified version of the Janus solutions found in \cite{Chiodaroli:2009yw} which were also put in correspondence with an interface CFT.\par
Similar to the Janus solutions, the approach outlined in this paper is relevant for the  supersymmetric wormhole solutions recently found in \cite{Astesiano:2023iql}. The existence of moduli, or scalar fields becoming moduli near the wormhole's neck, might be a crucial element in constructing such wormhole solutions.

We leave these projects to a future investigation.

\bigskip
\bigskip
\leftline{\bf Acknowledgements}
\smallskip
\noindent 

MT is grateful to Emanuel Malek, Tomas Ortin, Henning Samtleben and Thomas Van Riet for fruitful discussions. DA is grateful to Fri{\dh}rik Freyr Gautason for useful discussions.
DA is supported by the Icelandic Research Fund under grant 228952-052.

\newpage
\appendix
\section{Type IIB Supergravity on $T^4$: Geometric Characterization of the Scalar and Tensor Fields.} \label{AppendixA} In the solvable parametrization of the scalar manifold, the ten-dimensional dilaton $\phi$ and the radial moduli $\sigma_i$, $i=1,\dots, 4$, of $T^4$, parametrize the Cartan subalgebra of $\mathfrak{so}(5,5)$ while the remaining scalar fields are parameters of the shift generators $E_{\boldsymbol{\alpha}}$ corresponding to the positive roots $\boldsymbol{\alpha}$. The latter are conveniently expressed in terms of the simple roots which, in a suitable orthonormal basis $\boldsymbol{\epsilon}_i,\,\boldsymbol{\epsilon}_5$, are chosen to have the following form:
\begin{align}
\boldsymbol{\alpha}_1&=\boldsymbol{\epsilon}_1-\boldsymbol{\epsilon}_2\,,\,\,\boldsymbol{\alpha}_2=\boldsymbol{\epsilon}_2-\boldsymbol{\epsilon}_3\,,\,\,\boldsymbol{\alpha}_3=\boldsymbol{\epsilon}_3-\boldsymbol{\epsilon}_4\,,\,\,
\boldsymbol{\alpha}_4=\boldsymbol{\epsilon}_3+\boldsymbol{\epsilon}_4\,,\,\,\boldsymbol{\alpha}_5={\bf a}=-\frac{1}{2}\left(\sum_{i=1}^4\boldsymbol{\epsilon}_i\right)+\boldsymbol{\epsilon}_5\,.
\end{align}
The generators $t_\upalpha{}^{ij},\,t$ of the solvable Lie algebra generating the coset, which enter the definition \eqref{cosetr} of the coset representative correspond to the following shift-generators $E_{\boldsymbol{\alpha}}$:
\begin{align}
t_{\upalpha=1}{}^{ij}=E_{{\bf a}+\boldsymbol{\epsilon}_i+\boldsymbol{\epsilon}_j}\,,\,\,\,t_{\upalpha=2}{}^{ij}=E_{\boldsymbol{\epsilon}_i+\boldsymbol{\epsilon}_j}
\,,\,\,\,t=E_{{\bf a}+\boldsymbol{\epsilon}_1+\boldsymbol{\epsilon}_2+\boldsymbol{\epsilon}_3+\boldsymbol{\epsilon}_4}\,.
\end{align}
The six positive roots $\boldsymbol{\epsilon}_i-\boldsymbol{\epsilon}_j$, $i<j$, are the positive roots of the ${\rm GL}(4,\mathbb{R})$ group of the 4-torus and enter the definition of the vielbein matrix ${\bf E}\equiv (E_i{}^k)$. The latter can be written as $${\bf E}=\hat{\bf E}\cdot {\bf R}\,,$$
where
\begin{equation}
 \hat{\bf E}=\prod_{i<j} e^{\gamma_i{}^j\,E_{\boldsymbol{\epsilon}_i-\boldsymbol{\epsilon}_j}} \,\,,\,\,\,{\bf R}\equiv (\delta^i_j e^{\sigma_j})\,,
\end{equation}
and the generators $E_{\boldsymbol{\epsilon}_i-\boldsymbol{\epsilon}_j}$ are meant, in the above formula, as $4\times 4$ nilpotent matrices in the ${\bf 4}$ of ${\rm GL}(4,\mathbb{R})$. The dependence of the coset representative on the dilatonic scalars $\phi,\,\sigma_i$ is through the matrix 
\begin{equation}
    \mathbb{L}_{{\rm dil.}}\equiv \exp\left(\sum_{i=1}^4\sigma_i\,\left(H_{\boldsymbol{\epsilon}_i}+\frac{1}{2}\,H_{\boldsymbol{\epsilon}_5}\right)-\frac{\phi}{2}\,H_{{\bf a}}\right)\,.
\end{equation}
The dependence of the kinetic term of the non-dilatonic scalar ($\gamma_i{}^j,\,B^\upalpha{}_{ij},\,c_4$) on the dilatonic ones, in the Lagrangian, is through the characteristic exponential 
   $ e^{-2\,{\bf h}\cdot \boldsymbol{\alpha}}$,
where 
$${\bf h}\equiv \left(\sum_{i=1}^4\sigma_i\,\left({\boldsymbol{\epsilon}_i}+\frac{1}{2}\,{\boldsymbol{\epsilon}_5}\right)-\frac{\phi}{2}\,{{\bf a}}\right)\,,$$
 and $\boldsymbol{\alpha}$ is the positive root associated with the scalar.\par
 Similarly to the non-dilatonic scalars we can associate each of the ten 3-form field-strength $\mathbb{H}^M$ with weights ${\bf w}^M$ of the ${\bf 10}$ of $\mathfrak{so}(5,5)$ which,  the chosen conventions, read:
 \begin{equation}
 {\bf w}^M\,=\,({\bf w}^\upalpha,{\bf w}_{ij},\,{\bf w}_\upalpha)\,,
 \end{equation}
where:
\begin{align}
  {\bf w}^\upalpha&=\left( {\bf w}_0,\,-\boldsymbol{\epsilon}_5\right)\,,\,\, {\bf w}_{ij}={\bf w}_0+
 \boldsymbol{\epsilon}_i+\boldsymbol{\epsilon}_j\,,\,\,\,  {\bf w}_\upalpha=-{\bf w}^\upalpha\,,\,\,\,\,\,\,
 {\bf w}_0 \equiv -\frac{1}{2}\sum_{i=1}^4\boldsymbol{\epsilon}_i\,.
\end{align}
The dependence of the kinetic term of $\mathbb{H}^M$ on the dilatonic scalars in the Lagrangian is through the characteristic exponential: $e^{-2\,{\bf h}\cdot {\bf w}}$.\par
The $d_5,\,d_1$ charges are associated with the weights $\pm {\bf w}_0$.
The Cartan generator $\mathfrak{h}$ in \eqref{frakh} is defined by the component of ${\bf h}$ along ${\bf W}_0$ 
and is spanned by the scalar $g$:
$$e^g=e^{-{\bf h}\cdot {\bf w}_0}=g_s\,G^{\frac{1}{2}}\,.$$
The roots of ${\rm SO}(5,5)$ which are orthogonal to ${\bf w}_0$ define an ${\rm SO}(4,4)$ algebra representing the maximal choice of $G_0$, subgroup of $G$ which commutes with $\mathfrak{h}$. The positive roots of ${\rm SO}(4,4)$ and the corresponding non-dilatonic scalars are:
\begin{align}
    {\bf a}+\boldsymbol{\epsilon}_i+\boldsymbol{\epsilon}_j\,\,\leftrightarrow\,\,\,\,C_{ij}\,,\nonumber\\
   \boldsymbol{\epsilon}_i-\boldsymbol{\epsilon}_j \,\,,\,(i<j)\,\,\,\leftrightarrow\,\,\,\,\gamma_i{}^j\,,\nonumber
\end{align}
As for the dilatonic scalar fields, these are the components of ${\bf h}$ along the orthonormal basis ${\bf v}_i\equiv \frac{{\bf a}}{2}+ \boldsymbol{\epsilon}_i$, orthogonal to ${\bf w}_0$:
\begin{equation}
   \tilde{\sigma}_i= {\bf h}\cdot {\bf v}_i=\sigma_i-\frac{\phi}{4}\,.
\end{equation}
The scalars $\gamma_i{}^j,\,  \tilde{\sigma}_i$ define the components of the matrix $\tilde{G}_{ij}\equiv e^{-\frac{\phi}{2}}\,G_{ij}$.\par
The NS-NS charges $n_5,\,f_1$, associated with the NS 5-brane and the fundamental string, respectively, correspond to the weights $\mp\boldsymbol{\epsilon}_5$. 
The scalar $g$ parametrizing $\mathfrak{h}$ and entering eq. \eqref{NSg} is now given by 
$$e^g=e^{{\bf h}\cdot \boldsymbol{\epsilon}_5}=g_s^{-1}\,G^{\frac{1}{2}}=e^{-2\phi_6}\,,$$
where the six-dimensional dilaton is defined as follows: $e^{\phi_6}\equiv e^\phi\,{\rm det}\left(G^{(s)}_{ij}\right)^{-\frac{1}{4}}$ and $G^{(s)}_{ij}\equiv e^{\frac{\phi}{2}}\,G_{ij}$ is the torus metric in the string frame.
The dilatonic scalar fields in the ${\rm SO}(4,4)/{\rm SO}(4)\times {\rm SO}(4)$ coset, are given by the components of ${\bf h}$ along the orthonormal basis
$${\bf v}_i\equiv \boldsymbol{\epsilon}_i\,,\,\,\,i=1,\dots, 4\,.$$
They are:
$$\sigma_i^{(s)}\equiv  {\bf h}\cdot {\bf v}_i=\sigma_i+\frac{\phi}{4}\,,$$
and are the radial moduli in the string frame. On the other hand, the positive roots which are orthogonal to $\boldsymbol{\epsilon}_5$ are 
\begin{align}
  \boldsymbol{\epsilon}_i+\boldsymbol{\epsilon}_j\,\,\leftrightarrow\,\,\,\,B_{ij}\,,\nonumber\\
   \boldsymbol{\epsilon}_i-\boldsymbol{\epsilon}_j \,\,,\,(i<j)\,\,\,\leftrightarrow\,\,\,\,\gamma_i{}^j\,.\nonumber
\end{align}
This proves that the manifold ${\rm SO}(4,4)/{\rm SO}(4)\times {\rm SO}(4)$ is spanned by the moduli $G^{(s)}_{ij},\,B_{ij}$.
\section{The Type IIB Static Black String Solutions} \label{GeneralStringSolapp}
Let us study the most general ansatz for a static black string solution with ${\rm SO}(4)$ symmetry.  The corresponding metric has the following general form
\begin{gather}
    ds^2= A(r)^2(-dt^2+dx^2)+ A(r)^{-2} dr^2+ B(r)^2 \left[d\psi^2 + \sin^2\psi\left(d\theta^2+\sin^2\theta d\omega^2\right) \right],
\end{gather}
where $r,\psi,\theta,\omega$ are the polar coordinates of the  four-dimensional space transverse to the string world volume and the last three angles parameterize a 3-sphere $S^3$. The coordinates $x^{\hat{\mu}}$, $\hat{\mu}=0,\dots, 6$ naturally split into 
$$x^\mu=\{t,x,r\}\,\,,\,\,\,x^{\tilde{\mu}}=\{\psi,\theta,\omega\}\,,$$
where $\mu=0,1,2$ and $\tilde{\mu}=3,4,5$.\par
The Ricci tensor is 
\begin{align}
R_{tt}=&- R_{xx}= A^2 \left(A A''+ 3\frac{AA'B'}{B}+2 A'^2\right), \nonumber \\ R_{rr}=& \frac{1}{A^2B}\left[-3A \left(A'B'+AB''\right)-2B \left(A A''+A'^2\right)\right], \nonumber \\
R_{\psi\psi}=& \frac{R_{\theta\theta}}{\sin^2\psi}=\frac{R_{\omega\omega}}{\sin^2\psi \sin^2\theta}= -3 ABA'B'-A^2 \left[(BB''+2 B'^2)\right] \nonumber.
\end{align} 
We emphasize that this ansatz described the most general static black-string solution coupled to any number of scalar fields. The constant charge vector $\Gamma\equiv (\Gamma^M)$ is defined as follows:
$$\Gamma^M=\frac{1}{2\pi^2}\,\int_{S^3}\,\mathbb{H}^M\,,$$
and characterizes the solution. 
Let us consider the already introduced (\ref{AnsatzH}) ansatz for the tensor field
\begin{align}
    \mathbb{H}=- \xi(r) \Omega \mathcal{M} \Gamma dt \wedge dx \wedge dr+ \chi(\theta,\psi) \Gamma d\psi\wedge d\theta \wedge d\omega,
\end{align}
and this solves  the twisted self-duality condition, the Maxwell equations, and the Bianchi equations if 
\begin{gather}
    \xi(r)= \frac{A}{B^3}, \qquad \chi(\theta,\psi)= \sin^2 \psi \sin \theta.
\end{gather}
With this choice, the energy-momentum tensor reads
\begin{gather}
    T_{tt}=-T_{xx}=-A^4T_{rr}= \frac{A^2}{2B^6} V, \\
    T_{\psi\psi}=\frac{V}{2B^4}.
\end{gather}
On the inverse trace of Einstein's equation
\begin{gather}
    R_{\hat{\mu}\hat{\nu}}= T^{(H)}_{\hat{\mu}\hat{\nu}}+ \frac{1}{2} \mathcal{G}_{ts}\partial_{\hat{\mu}}\phi^t\partial_{\hat{\nu}}\phi^s,
\end{gather}
we impose the consistency requirement $\phi=\phi(r)$ and the equations explicitly read
\begin{align}
    R^{t}_{t}= R^{x}_{x}= - \frac{V}{2B^6}, \\
    R^{r}_{r}= - \frac{V}{2B^6}+  \frac{1}{2} \mathcal{G}_{ts}\partial_{r}\phi^t\partial^{r}\phi^s, \\
     R^{\psi}_{\psi}= R^{\theta}_{\theta}= R^{\omega}_{\omega}= \frac{V}{2B^6}.
\end{align}
These equations force some consistency relations such that 
\begin{align}
     R^{t}_{t}=R^{x}_{x}, \qquad  R^{\psi}_{\psi}= R^{\theta}_{\theta}= R^{\omega}_{\omega},
\end{align}
which are already verified, but there is another condition that we still have to impose, i.e.
\begin{align}
   R^{t}_{t}+ R^{\psi}_{\psi}=0 \rightarrow \frac{1}{2} \left(A^2B^2\right)''+ (AB)'^2-2=0.
\end{align}
Posing $AB=\pm \sqrt{u}$ this condition becomes
\begin{gather}
    \frac{d^2u}{dr^2}+ \frac{1}{2u} \left(\frac{du}{dr}\right)^2-4=0.
\end{gather}
To solve this we define
\begin{align}
    \Omega(u)= \frac{du}{dr}, \qquad \dot\Omega= \frac{d}{du}\Omega,  \label{COND}
\end{align}
and we get the Bernoulli equation
\begin{align}
    \Omega \dot{\Omega}+ \frac{\Omega^2}{2u} -4=0,
\end{align}
with solutions
\begin{align}
    \Omega= \pm \sqrt{\frac{a+4u^2}{u}}.
\end{align}
We choose the "+" sign and, from eq. $(\ref{COND})$, we have
\begin{align}
    r-c= \int \frac{du}{\sqrt{\frac{a+4u^2}{u}}}= \frac{2u \sqrt{\frac{4u^2}{a}+1}}{3 \sqrt{\frac{a}{u}+4u}}  {}_2\text{F}_1 \left(\frac{1}{2},\frac{3}{4},\frac{7}{4},\frac{-4u^2}{a}\right), \qquad u= A(r)^2 B(r)^2.
\end{align}
The extremal solutions are recovered with $a\rightarrow 0$. In fact, this gives
\begin{align}
    (r-c)^2 = A(r)^2 B(r)^2,
\end{align}
from which we can recognize the structure of the double coincident horizon at $r = c$.\\
At radial infinity, $r$ and $u$ go to $+\infty$, while, near the horizon $r\sim c$ and $u\sim 0$. In these parameterizations, however, the inner and outer horizons do not appear. 
Now with a little abuse of notation we regard $A$ and $B$ as functions of $u$ and we trade $B$ with $u$ in the metric:
\begin{align}
    ds^2=A(u)^2 \left(-dt^2+dx^2\right)+ u\, A(u)^{-2}\left(\frac{du^2}{a+4u^2} +  d\Omega^2_{S^3} \right).
\end{align}
The affine parameter $\tau$ can be defined by the differential equation
\begin{align}
    \frac{d\tau}{du}= \frac{1}{u\sqrt{a+4u^2}},
\end{align}
which can be solved as follows:
\begin{equation}
    u(\tau)=-\frac{\sqrt{a}}{2\sinh(\sqrt{a}\tau)}\,.
\end{equation}
Notice that we have chosen the sign of $ u(\tau)$ so that, since $u\ge 0$, $\tau$ be non-positive. At radial infinity ($u\rightarrow \infty$) $\tau\rightarrow 0^-$and near the horizon ($u\rightarrow 0$)  $\tau\rightarrow -\infty$.\\
In terms of the affine parameter and defining $A(u)= e^{U(\tau)}$ the non-extremal metric is
\begin{align}
    ds^2=e^{2U}(-dt^2+dx^2)+ \frac{1}{2}  \frac{\sqrt{a}}{\sinh(\sqrt{a}\tau)} e^{-2U(\tau)}\left(\frac{1}{4} \frac{a}{\sinh^2(\sqrt{a}\tau)} d\tau^2+ d\Omega^2_{S^3} \right).
\end{align}
Regularity requires $a>0$.
Einstein's equations and the scalar fields equations are
\begin{align}
    \ddot{U}=& e^{4U}\frac{V}{2}, \label{Einst1}\\
     \ddot{\phi}^s+ \tilde{\Gamma}^s_{tu} \dot{\phi}^t \dot{\phi}^u=&e^{4U} \mathcal{G}^{st}\partial_s V, \label{Einst2}\\
     \frac{3a}{8}=&  \dot{U}^2- e^{4U}\frac{V}{4}+ \frac{1}{8}\mathcal{G}_{ts}\dot \phi^t \dot\phi^s \label{Einst3},
\end{align}
where the dots are the derivatives with respect to the affine parameter $\tau$. 
The first two equations can be deduced from the effective action
\begin{align}
    S_{\text{eff}}= \int d\tau\,\left( \dot{U}^2 + \frac{1}{8} G_{su} \dot{\phi}^s \dot{\phi}^u + e^{4U} \frac{V}{4}\right),
\end{align}
while the third one can be interpreted as a Hamiltonian constraint, with energy $\frac{3a}{8}$.\\
The near horizon limit can be obtained by taking $\tau \rightarrow -\infty$ and the metric reads
\begin{align}
    ds^2=e^{2U}(-dt^2+dx^2)+  \sqrt{a} e^{\sqrt{a}\tau} e^{-2U(\tau)} \left(a e^{-2\sqrt{a}\tau} d\tau^2+ d\Omega^2_{S^3} \right).
\end{align}
To have a finite horizon area, we must require the following behavior
\begin{align}
   e^{-2U} \sim \left( \frac{A_H}{2\pi^2}\right)^{2/3} \frac{1}{\sqrt{a}e^{\sqrt{a}\tau}}, \text{as} \quad \tau \rightarrow \infty. 
\end{align}
The two horizons can be described by changing the radial variable into a new one $\rho$ defined through the relation:
\begin{equation}
    \sinh^2(\sqrt{a}\tau)=\frac{a}{(\rho-\rho_0)^2-a}=\frac{a}{4u^2}\,,
\end{equation}
which can be solved in the two variables as follows:
\begin{equation}
    \rho=\rho_0-\sqrt{a}\,\coth(\sqrt{a}\tau)\,\,,\,\,\,\tau=\frac{1}{2\sqrt{a}}\,\log\left(\frac{\rho-\rho_+}{\rho-\rho_-}\right)\,,
\end{equation}
where we have denoted by $\rho_\pm$ the radial location of the two horizons defined as:
\begin{equation}
    \rho_\pm\equiv \rho_0\pm \sqrt{a}\,.
\end{equation}
Therefore, the metric near the horizon reads
\begin{align}
    ds^2= \left( \frac{A_H}{2\pi^2}\right)^{-2/3} \sqrt{a} e^{\sqrt{a}\tau} (-dt^2+dx^2)+ \left( \frac{A_H}{2\pi^2}\right)^{2/3} \left(  a e^{2 \sqrt{a}\tau} d\tau^2+ d\Omega^2_{\text{S}^3}\right).
\end{align}
The existence of a timelike killing vector $\xi=\partial_t$, guarantees the existence of the Komar mass for this class of solutions, which is given by 
\begin{align}
   M= \frac{c^2}{8 \pi G} \int_{S^3_{\infty}} \sqrt{g}|_{\theta,\phi,\omega} \epsilon_{\theta \phi \omega \mu \nu} \nabla^{\mu} \xi^{\nu}=
    \frac{\pi c^2}{4 G} \lim_{\tau \rightarrow 0^{-}} \dot{U}.
\end{align}
This allows us to fix the following conditions at infinity
\begin{align}
    U(0)=0, \quad \dot{U}(0)= M \frac{4 G}{\pi c^2}.
\end{align}
\subsection{Attractor mechanism for extremal solutions}
If $a=0$ we have
\begin{align}
    \tau=& -\frac{1}{2u}\,\,,\,\,\,\,\rho-\rho_0=-\frac{1}{\tau}=2\,r^2\,,
\end{align}
and the two horizons coincide $\rho_\pm=\rho_0$.
The metric reads
\begin{align}
     ds^2=e^{2U} \left(-dt^2+dx^2\right)+ \frac{1}{(-2\tau)}\, e^{-2U}\left(\frac{d\tau^2}{4 \tau ^2} +  d\Omega^2_{S^3} \right).
\end{align}
The horizon is at again at $u \rightarrow 0$ and $\tau \rightarrow -\infty$ and the near horizon behaviour of the function $e^{-U}$ is now given by
\begin{align}
    e^{-2U} =\left(\frac{A_H}{2\pi^2}\right)^{2/3} (-2 \tau), \text{as} \quad \tau \rightarrow \infty. \label{RegCondExt}
\end{align}
Since the area of $S_3$ is $A_H=2 \pi^2 L^3$, we can also replace $A_H$ with the radius of the three-sphere $L$.
The coordinate which defines the proper spatial distance from the horizon is
\begin{align}
  d\zeta=  \frac{e^{-U}}{\sqrt{(-8\tau^3)}}\, d\tau= -\frac{L}{2} d\log(- \tau).
\end{align} 
Thanks to the regularity condition in eq.(\ref{RegCondExt}) and fixing a not-useful constant we find
\begin{align}
    \zeta=-\frac{L}{2} \log(-\tau)+{\rm const.},
\end{align}
while the position of the horizon is now at $\zeta \rightarrow -\infty$. Again, we require the scalar fields to be regular at the horizon
\begin{gather}
    \lim_{\zeta \rightarrow -\infty} \phi^s(\zeta)= \phi^s_* < \infty. \label{RegCond}
\end{gather}
If the functions $\phi^s$ are uniformly continuous this request necessarily implies that the derivatives 
\begin{gather}
   \lim_{\zeta\rightarrow -\infty} \frac{d^k}{d\zeta^k}\phi^s =0.
\end{gather}
This implies
\begin{align}
    \lim_{\tau \rightarrow -\infty} \tau \frac{d}{d\tau}\phi^s=\lim_{\tau \rightarrow -\infty} \tau^2 \frac{d^2}{d\tau^2}\phi^s=...=0. \label{RegRule}
\end{align}
The scalar field equation (\ref{Einst2}) can be rewritten as
\begin{gather}
    \tau^2 \partial^2_{\tau} \phi^s+ \tilde{\Gamma}^s_{uv} \left(\tau \partial_{\tau}\phi^{u}\right) \left(\tau \partial_{\tau}\phi^{v} \right)= \mathcal{G}^{st}\partial_t V\,L^{-6}.
\end{gather}
Taking the limit at the near horizon and applying the rule for regularity (\ref{RegRule}) we get the  \emph{attractor mechanism}\footnote{The attractor mechanism was first found for $D=4$ asymptotically-flat, extremal  black holes in \cite{Ferrara:1996um,Ferrara:1997tw}.} equation
\begin{gather}
    \lim_{\phi^s \rightarrow \phi^s_*} \partial_t V=0.
\end{gather}
The scalar fields which $V$ depends on are attracted towards fixed value extremizing the potential, while the others are called flat directions. Therefore, the attractor mechanism forces the effective potential $V$ to have an extremum at the horizon, namely for $\phi^s=\phi^s_*$:
\begin{gather}
  \Gamma^T \partial_s \mathcal{M} \Gamma|_{\text{horizon}}=0\,. \end{gather} 
We can also find the explicit critical value of the black hole potential at the horizon. To do this, we note that the function $U$ near the horizon is explicitly given by
\begin{align}
    U=-\frac{1}{2} \log\left({L^2 (-2\tau)}\right).
\end{align}
If we insert this result inside eq.(\ref{Einst1})-(\ref{Einst3}) we get that the value of the black hole potential $V$ at the horizon is fixed to be
\begin{align}
    V_{\text{horizon}}= V_*=4\,L^4.
\end{align}
Restoring the coordinate $r$, and setting $c=0$, the metric at the horizon takes the form
\begin{align}
    ds^2= \frac{r^2}{r^2_H} (-dt^2+dx^2)+ \frac{L^2}{r^2}dr^2+L^2 d\Omega_{S^3},
\end{align}
which describes an ${\rm AdS}_3\times S^3$ geometry.
\section{ String solutions in the double commuting $\frac{{\rm SL}(2,\mathbb{R})}{{\rm SO}(2)}$ truncation} \label{Stringsolutionsapp}
We can now write the solutions of the equations (\ref{Einst1})-(\ref{Einst3}) by minimizing the potential $V$, as discussed in appendix \ref{CriticalPointsPot}. After this minimization, to solve the equations we just need the moduli to move along geodesics on the target space. Here we discuss the case 
of two commuting ${\rm SL}(2,\mathbb{
R})$ of the scalar manifold, for which the potential and the metric reads
\begin{align}
   d\tilde s=& d\phi^2+d\psi^2+d\varphi^2+ e^{\varphi-\sqrt{2} \psi+\phi} (d C_{12})^2+e^{\varphi+\sqrt{2} \psi+\phi}(d C_{34})^2, \label{metricSL2} \\
    V=& \frac{e^\varphi}{2} \left( d_1^2e^{-\phi} + d_5^2 e^{\phi-2\varphi}\right),
\end{align}
where $B^1_{12}=C_{12}$ and $B^1_{34}=C_{34}$.
From the perspective of the 10D, the metric in Einstein's frame is
\begin{align}
& ds^2_{10} = e^{2\alpha\varphi} ds^2_{6} + e^{2\beta\varphi}\left(e^{2\gamma\psi} d \theta_1^2 + e^{2\gamma\psi} d \theta_2^2+ e^{-2\gamma\psi} d \theta_3^2 + e^{-2\gamma\psi} d \theta_4^2\right)\,,\\
& C_2 = \tilde{C}_2+  C_{12} d\theta_1\wedge d\theta_2 + C_{34} d\theta_3\wedge d\theta_4\,,
\end{align}
where $\alpha=1/4=-\beta$ and $\gamma^2=1/8$.\\
In this case the einstein equations are, directly from eq.(\ref{Einst1})-(\ref{Einst3})
\begin{align}
    \ddot{U}(\tau )&-\frac{1}{4} e^{\varphi(\tau)+4 U(\tau )} \left( d_1^2 e^{-\phi
   (\tau )}+ d_5^2 e^{\phi (\tau )-2 \varphi (\tau )}\right)=0, \nonumber\\
   \ddot{\phi}(\tau )&+ \frac{1}{2}\,e^{\varphi (\tau )+4 U(\tau )} \left( -d_5^2
   e^{\phi(\tau )-2 \varphi (\tau)}+ d_1^2 e^{\phi(\tau)}\right)-\frac{1}{2} e^{\varphi(\tau)-\sqrt{2}\psi(\tau)+\phi(\tau)}(\dot{C}_{12}(\tau)^2+e^{2\sqrt{2}\psi(\tau)}\dot{C}_{34}(\tau)^2)=0,\nonumber\\
    \ddot{\varphi}(\tau )&+\frac{1}{2}\, e^{4 U(\tau )} \left( d_5^2
   e^{\phi(\tau )- \varphi (\tau)}- d_1^2 e^{+\varphi(\tau)-\phi(\tau)}\right)-\frac{1}{2} e^{\varphi(\tau)-\sqrt{2}\psi(\tau)+\phi(\tau)}(\dot{C}_{12}(\tau)^2+e^{2\sqrt{2}\psi(\tau)}\dot{C}_{34}(\tau)^2)=0,\nonumber\\
   \ddot{\psi}(\tau)&+\frac{1}{\sqrt 2} e^{\varphi(\tau)-\sqrt{2}\psi(\tau)+\phi(\tau)}(\dot{C}_{12}(\tau)^2-e^{2\sqrt{2}\psi(\tau)}\dot{C}_{34}(\tau)^2)=0, \nonumber\\
   \ddot{C}_{12}(\tau)&+\dot{C}_{12}(\tau)\left(\dot{\varphi}(\tau)-\sqrt{2} \dot{\psi}(\tau)+\dot{\phi}(\tau)\right)=0,\nonumber\\
   \ddot{C}_{34}(\tau)&+\dot{C}_{34}(\tau)\left(\dot{\varphi}(\tau)+\sqrt{2} \dot{\psi}(\tau)+\dot{\phi}(\tau)\right)=0.
\end{align}
In order to separate the moduli from the other scalars we introduce the function
\begin{align}
    \varphi=\frac{h(\tau)-g(\tau)}{2}, \quad \phi=\frac{h(\tau)+g(\tau)}{2},
\end{align}
therefore the system totally splits into two different sets of equations. This is due to the diagonal form of the metric, see eq.(\ref{metricSL2}), in this particular truncation. The first equations are for the scalars which are not modules
\begin{align}
\ddot{U}(\tau )&-\frac{1}{4} e^{4 U(\tau )-g(\tau )} \left(d_1^2+d_5^2 e^{2 g(\tau )}\right)=0, \\
  \ddot{g}(\tau )&+   e^{4 U(\tau )-g(\tau )} \left(d_1^2-d_5^2 e^{2 g(\tau )}\right)=0,
\end{align}
and the second system is for the moduli
\begin{align}
\ddot{h}(\tau )&-e^{h(\tau )-\sqrt{2} \psi(\tau )} \left(\dot{C}_{12}(\tau )^2+e^{2 \sqrt{2}
   \psi(\tau )} \dot{C}_{34}(\tau )^2\right)=0\label{moduli1},\\
    \ddot{\psi}(\tau)&+\frac{1}{\sqrt 2} e^{h(\tau)-\sqrt{2}\psi(\tau)}\left(\dot{C}_{12}(\tau)^2-e^{2\sqrt{2}\psi(\tau)}\dot{C}_{34}(\tau)^2\right)=0,\label{moduli2}\\
   \ddot{C}_{12}(\tau)&+\dot{C}_{12}(\tau)\left(\dot{h}(\tau)-\sqrt{2} \dot{\psi}(\tau)\right)=0,\label{moduli3}\\
   \ddot{C}_{34}(\tau)&+\dot{C}_{34}(\tau)\left(\dot{h}(\tau)+\sqrt{2} \dot{\psi}(\tau)\right)=0\label{moduli4}.
\end{align}

\subsection{D1-D5 Solution \label{D1-D5}}
When the moduli are equal to zero, to solve the first system we can define the superpotential to be
\begin{align}
    W=e^{-\frac{g(\tau )}{2}} \left(d_1+d_5 e^{g(\tau )}\right).
\end{align}
Thanks to this definition we can prove that the system can be cast into a first-order system
\begin{align}
    \dot{g}(\tau )-2\, e^{2 U(\tau )} \partial_{g(\tau)} W&=0,
    \\
    \dot{U}(\tau )-\frac{W e^{2 U(\tau )}}{4}&=0.
\end{align}
Now it is easy to see that the solution is given by
\begin{align}
    U(\tau)&= -\frac{1}{4} \log[{(1-d_1 \tau)(1-d_5 \tau)}],\\
    -\varphi(\tau)&=\phi(\tau)= \frac{1}{2} \log\left[{\frac{1-d_1 \tau}{1-d_5 \tau}}\right],\\
    \psi(\tau)&=c_{12}(\tau)=c_{34}(\tau)=0,
\end{align}
for which the metric reads 
\begin{align}
    ds^2= \frac{(-dt^2+dx^2)}{\sqrt{(-d_1  \tau +1) (-d_5 \tau +1)}}+\frac{{\sqrt{(-d_1  \tau +1) (-d_5 \tau +1)}}}{(-2 \tau)}\left(\frac{d\tau^2}{4 \tau ^2} +  d\Omega^2_{S^3} \right).
\end{align}
We can now define the usual radial coordinate $\tau= (-2 r^2)^{-1}$ and we end up with the usual form of the D1-D5 system
    \begin{align}
    ds^2 =& (Z_1 Z_5)^{-\frac{1}{2}} (-dt^2+dx^2)+ (Z_1 Z_5)^{\frac{1}{2}} (dx^i dx^i), \\
    dx^i dx^i =& dr^2 + r^2 d\Omega^2_3, \qquad Z_1=1+ \frac{Q_1}{r^2},\qquad Z_5= 1+\frac{Q_5}{r^2},
\end{align}
where $Q_1=d_1/2$ and $Q_5=d_5/2$. At the horizon the $Z$ functions become
\begin{align}
   Z_1= \frac{Q_1}{r^2},\qquad Z_5= \frac{Q_5}{r^2},
\end{align}
and the physical distance from the horizon is
\begin{gather}
    \rho= \log\left( \frac{r}{\ell}\right).
\end{gather}
This background is ${\rm AdS}_3 \times S^3$, which is the throat of the D1-D5 system.
     Now, the scalars are attracted towards configurations that extremizes this potential, called critical points \begin{gather}
    \partial_\phi V=0 \rightarrow V=V_*=d_1\,d_5.
\end{gather}
Since we are in the near-horizon region where this critical point is reached by the scalar fields, we have $V=V_*$ and the energy-momentum tensor for the scalars is zero $T^{(S)}_{\hat{\mu}\hat{\nu}}=0$. 
The near-horizon metric is conveniently written in the following form:
\begin{gather}
    ds^2= \frac{r^2}{L^2}(-dt^2+dx^2)+ \frac{L^2}{r^2} dr^2+ L^2 \left[d\psi^2 + \sin^2\psi\left(d\theta^2+\sin^2\theta d\omega^2\right) \right],
\end{gather}
where $L^2=\frac{\sqrt{d_1 d_5}}{2}$.
Einstein's equations boil down to \begin{gather}
    G_{\mu}^{\nu}=-\frac{1}{2} V_{0} \delta_{\mu}^{\nu}= T^{(H)\nu}_{\,\,\mu}, \quad
G_{\bar{\mu}}^{\bar{\nu}}=\frac{1}{2} V_{0} \delta_{\bar{\mu}}^{\bar{\nu}}=T^{(H)\bar{\nu}}_{\,\,\bar{\mu}},
\end{gather}
solved by the condition $L^2=\sqrt{Q_1 Q_5}$.
\subsection{D1-D5 solution with non-trivial moduli}
The equations for the moduli (\ref{moduli1})-(\ref{moduli4}) are consistent with all of them to be zero, this was the matter of the last section \ref{D1-D5} which describes the D1-D5 system. In this subsection, we consider a solution in which some of the moduli fields are allowed to evolve in the radial variable $\tau$. The resulting solution, however, is not related to the general construction considered in this work, in which the evolution of the moduli is on one of the boundary coordinates instead.  \\
We can still keep part of the solution we wrote before 
\begin{align}
    U(\tau)&= -\frac{1}{4} \log[{(1-d_1 \tau)(1-d_5 \tau)}],\\
    g(\tau)&= \log\left[{\frac{1-d_1 \tau}{1-d_5 \tau}}\right],
\end{align}
while the moduli remain to be fixed now. If we now introduce
\begin{align}
    s(\tau)= h(\tau)+\sqrt{2}\psi(\tau), \quad t(\tau)=h(\tau)-\sqrt{2}\psi(\tau),
\end{align}
 the equations decouple into two independent systems
\begin{align}
\ddot{t}(\tau )&-2 e^{t(\tau )} \dot{c}_{12}(\tau )^2=0,\quad
\ddot{c}_{12}(\tau)+\dot{c}_{12}(\tau )\dot{t}(\tau)=0,\\
\ddot{s}(\tau )&-2 e^{s(\tau )} \dot{c}_{34}(\tau )^2=0,\quad
\ddot{c}_{34}(\tau)+\dot{c}_{34}(\tau )\dot{s}(\tau)=0.
\end{align}
We can directly integrate $c_{12}$ and $c_{34}$ as
\begin{align}
    c_{12}(\tau)= c_1 \int e^{-t(\tau)} d\tau, \quad
c_{34}(\tau)= c_3 \int e^{-s(\tau )} d\tau,
\end{align}
while the remaining equations are
\begin{align}
    \quad \ddot{t}(\tau )=2 c_1^2 e^{-t(\tau )}, \quad \ddot{s}(\tau )=2 c_3^2 e^{-s(\tau )},
\end{align}
where $c_1$ and $c_3$ are constants. The whole system can now be integrated and the solution reads
\begin{align}
t&=\log \left[-2 c_1^2 (\cosh (\tau )-1)\right], \quad c_{12}=\frac{\coth \left(\frac{\sqrt{6a} }{2} \tau\right)}{2 c_1},\nonumber\\
s&=\log \left[-2 c_3^2 (\cosh (\tau )-1)\right],\quad c_{34}=\frac{\coth \left(\frac{\sqrt{6a} }{2} \tau\right)}{2 c_3} \nonumber.
\end{align}
In terms of the field appearing in the truncation given by (\ref{metricSL2}), the solution is
\begin{align}
    \varphi&=\frac{1}{4} \left[-2 \log \left(\frac{1-d_1 \tau }{1-d_5 \tau }\right)+\log \left(4 c_1{}^2 c_3{}^2 (\cosh (\sqrt{6a}  \tau )-1)^2\right)\right] \nonumber, \\
   \phi&=\frac{1}{4} \left[2 \log \left(\frac{1-d_1 \tau }{1-d_5 \tau }\right)+\log \left(4 c_1{}^2 c_3{}^2 (\cosh ( \sqrt{6a}\tau )-1)^2\right)\right] \nonumber,\\
   \psi&=\frac{\log \left(-\frac{c_3}{c_1}\right)}{\sqrt{2}},\quad
   c_{12}=\frac{\coth \left(\frac{\sqrt{6a} }{2} \tau\right)}{2 c_1},\quad
   c_{34}=\frac{\coth \left(\frac{\sqrt{6a} }{2} \tau\right)}{2 c_3} \nonumber.
\end{align}
This solution is non-extremal. We can directly build other solutions using the S-duality transformation given by 
eq.(\ref{Sdual1})-(\ref{Sdual3}).

\section{Extremization of the effective scalar potential} \label{CriticalPointsPot}
In this section, we wish to discuss the extremization of $V(\phi,\Gamma)$. To this end, we first need to explicitly construct $\mathcal{M}$ in terms of the scalar fields by using the solvable parametrization of the coset manifold. 
Using the explicit dependence of $\mathcal{M}_{MN}$ on the dimensionally reduced Type IIB fields we can derive the moduli space of the ${\rm AdS}_3\times {\rm S}^3$ background.
We can derive the expression for the potential $V$ by first writing the charge vector $\Gamma^M$ in components:
\begin{equation}
\label{ChargesDef}
\Gamma^M = \left(
             \begin{array}{ccc}
               n{}^\alpha & D_{ij} & n_{\alpha} \\
             \end{array}
           \right)\,,
\end{equation}
where $n^\alpha=(d_5,n_5)$, $n_{\alpha}=(d_1,f_1)$, being $d_1,\,d_5$ the D1, D5 charges and $f_1,n_5$ the charges of the fundamental string and of the NS 5-brane, respectively, while $D_{ij}$ are the D3-brane charges. We then restrict ourselves to the D1-D5 charges and obtain:
\begin{align}\label{Pot}
2\,V\,&=\,\Gamma^M \mathcal{M}_{MN} \Gamma^N = \frac{G^{-\frac{1}{2}}}{g_s }\left(d_1-\frac{1}{8}\epsilon^{ijkl}B_{ij}B_{kl}d_5\right)^2+\\ \nonumber
&+g_s G^{-\frac{1}{2}}\left[\left(d_1-\frac{1}{8}\epsilon^{ijkl}B_{ij}B_{kl}d_5\right) C_{(0)}-\left(c-\frac{1}{8}\epsilon^{ijkl}B_{ij}C_{kl}\right)d_5\right]^2+\\ \nonumber
&+ g_s  G^{\frac{1}{2}}\left(\frac{1}{2 g_s } B_{ij}B_{kl}  G^{ik}G^{jl}+1\right)d_5^2\,.
\end{align}
The mass formula for the D1D5 system given in \cite{Larsen:1999uk} is obtained by adding an invariant, constant shift-term to the potential $V$, that is $\Gamma^M  \Omega_{MN} \Gamma^N/2$. The minimizing conditions over the moduli space are the following:
\begin{align}
c-\frac{1}{8}\epsilon^{ijkl}B_{ij}C_{kl}  &= g_s G^{\frac{1}{2}} C_{(0)}\,,\label{cond1}\\
g_s G^{\frac{1}{2}} + \frac{1}{8}\epsilon^{ijkl}B_{ij}B_{kl}&=\frac{d_1}{d_5}\,,\label{cond2}\\
 G^{\frac{1}{2}} B_{ij}  G^{ik}G^{jl} &= \frac{1}{2}\,B_{ij} \epsilon^{ijkl} \label{cond3}\,.
\end{align}
With these conditions the potential at the minimum is
\begin{equation}
V_*=d_1 d_5\,,
\end{equation}
which, after the mentioned shift, becomes exactly double.
The initial space for the scalar manifold is ${\rm SO}(5,5)/\left({\rm SO}(5)\times {\rm SO}(5)\right)$; due to the attractor mechanism, $5$ of these scalars are fixed at the minimum of the potential. These conditions are explicitly given in ($\ref{cond1}$), ($\ref{cond2}$) and ($\ref{cond3}$) .\\
Then, the scalar manifold reduces to ${\rm SO}(4,5)/\left({\rm SO}(4)\times {\rm SO}(5)\right)$ as required by CFT duality (see Cecotti) . To see this, let us show that ${\rm SO}(4,5)$ is the little group of the charge vector $\Gamma^M$ when only the D1-D5 charges are switched on:
\begin{equation}
\Gamma^M=(d_5,0,0,0,0,0,0,0,d_1,0)\,.
\end{equation}
Supersymmetry requires $d_1\,d_5>0$. Let us perform the following duality transformation:
\begin{equation}
\Gamma^M\,\rightarrow\,\,\,\Gamma^{\prime M}=\mathcal{O}^M{}_N\,\Gamma^N\,,
\end{equation}
where
\begin{equation}
\mathcal{O}^M{}_N={\rm diag}\left(\sqrt{\frac{d_1}{d_5}},1,1,1,1,1,1,1,\sqrt{\frac{d_5}{d_1}},1\right)\in {\rm SO}(5,5)\,.
\end{equation}
The new charge vector reads:
\begin{equation}
\Gamma^{\prime M}=\sqrt{d_1d_5}(1,0,0,0,0,0,0,0,1,0)\,.
\end{equation}
Changing the basis of the representation space by a Cayley transformation
$$\mathcal{C}^M{}_N=\frac{1}{\sqrt{2}}\,\left(\begin{matrix}{\bf 1} & {\bf 1}\cr {\bf 1} & -{\bf 1}\end{matrix}\right)\,,$$
the invariant matrix becomes:
\begin{equation}
\Omega\rightarrow \mathcal{C}^t\,\Omega\,\mathcal{C}={\rm diag}\left(1,1,1,1,1,-1,-1,-1,-1,-1\right)
\end{equation}
and the charge vector acquires the following form:
\begin{equation}
\Gamma\rightarrow \mathcal{C}\,\Gamma=\sqrt{2\,d_1d_5}\,{\rm diag}\left(1,0,0,0,0,0,0,0,0,0\right)\,.
\end{equation}
From the above form of the charge vector, it is straightforward to identify the ${\rm SO}(4,5)$ subgroup of ${\rm SO}(5,5)$ which leaves it invariant. It consists of those matrices which have a trivial action of the first entry of the vector.\\In the non-BPS case in which $d_1 d_5<0$ (e.g. $D1$-anti-$D5$ system), the same transformation yields the vector:
\begin{equation}
 \mathcal{C}\,\Gamma=\sqrt{2\,|d_1d_5|}\,{\rm diag}\left(0,0,0,0,0,0,0,0,1,0\right)\,.
\end{equation}
The stabilizer being still ${\rm SO}(4,5)$. This orbit, as opposed to the BPS one, is characterized by the invariant property $\Gamma^t\Omega \Gamma <0$.

\bibliography{refs}

\providecommand{\href}[2]{#2}\begingroup\raggedright\begin{thebibliography}{10}

\bibitem{Aspinwall:1996mn}
P.~S. Aspinwall, {\it {K3 surfaces and string duality}},  in {\em {Theoretical
  Advanced Study Institute in Elementary Particle Physics (TASI 96): Fields,
  Strings, and Duality}}, pp.~421--540, 11, 1996.
\newblock \href{http://arxiv.org/abs/hep-th/9611137}{{\tt hep-th/9611137}}.

\bibitem{David:2002wn}
J.~R. David, G.~Mandal, and S.~R. Wadia, {\it {Microscopic formulation of black
  holes in string theory}},  {\em Phys. Rept.} {\bf 369} (2002) 549--686,
  [\href{http://arxiv.org/abs/hep-th/0203048}{{\tt hep-th/0203048}}].

\bibitem{Avery:2010qw}
S.~G. Avery, {\it {Using the D1D5 CFT to Understand Black Holes}},  other
  thesis, 12, 2010.

\bibitem{Kachru:1998ys}
S.~Kachru and E.~Silverstein, {\it {4-D conformal theories and strings on
  orbifolds}},  {\em Phys. Rev. Lett.} {\bf 80} (1998) 4855--4858,
  [\href{http://arxiv.org/abs/hep-th/9802183}{{\tt hep-th/9802183}}].

\bibitem{Katmadas:2018ksp}
S.~Katmadas, D.~Ruggeri, M.~Trigiante, and T.~Van~Riet, {\it {The holographic
  dual to supergravity instantons in $\rm AdS_5\times S^5/\mathbb{Z}_k$}},
  {\em JHEP} {\bf 10} (2019) 205, [\href{http://arxiv.org/abs/1812.05986}{{\tt
  arXiv:1812.05986}}].

\bibitem{Corrado:2002wx}
R.~Corrado, M.~Gunaydin, N.~P. Warner, and M.~Zagermann, {\it {Orbifolds and
  flows from gauged supergravity}},  {\em Phys. Rev.} {\bf D65} (2002) 125024,
  [\href{http://arxiv.org/abs/hep-th/0203057}{{\tt hep-th/0203057}}].

\bibitem{Louis:2015dca}
J.~Louis, H.~Triendl, and M.~Zagermann, {\it {$ \mathcal{N}=4 $ supersymmetric
  AdS$_{5}$ vacua and their moduli spaces}},  {\em JHEP} {\bf 10} (2015) 083,
  [\href{http://arxiv.org/abs/1507.01623}{{\tt arXiv:1507.01623}}].

\bibitem{Hertog:2017owm}
T.~Hertog, M.~Trigiante, and T.~Van~Riet, {\it {Axion Wormholes in AdS
  Compactifications}},  {\em JHEP} {\bf 06} (2017) 067,
  [\href{http://arxiv.org/abs/1702.04622}{{\tt arXiv:1702.04622}}].

\bibitem{Ruggeri:2017grz}
D.~Ruggeri, M.~Trigiante, and T.~Van~Riet, {\it {Instantons from geodesics in
  AdS moduli spaces}},  {\em JHEP} {\bf 03} (2018) 091,
  [\href{http://arxiv.org/abs/1712.06081}{{\tt arXiv:1712.06081}}].

\bibitem{Astesiano:2022qba}
D.~Astesiano, D.~Ruggeri, M.~Trigiante, and T.~Van~Riet, {\it {Instantons and
  no wormholes in $AdS_3\times S^3 \times CY_2$}},  {\em Phys. Rev. D} {\bf
  105} (2022), no.~8 086022, [\href{http://arxiv.org/abs/2201.11694}{{\tt
  arXiv:2201.11694}}].

\bibitem{Guarino:2019oct}
A.~Guarino and C.~Sterckx, {\it {S-folds and (non-)supersymmetric Janus
  solutions}},  {\em JHEP} {\bf 12} (2019) 113,
  [\href{http://arxiv.org/abs/1907.04177}{{\tt arXiv:1907.04177}}].

\bibitem{Inverso:2016eet}
G.~Inverso, H.~Samtleben, and M.~Trigiante, {\it {Type II supergravity origin
  of dyonic gaugings}},  {\em Phys. Rev. D} {\bf 95} (2017), no.~6 066020,
  [\href{http://arxiv.org/abs/1612.05123}{{\tt arXiv:1612.05123}}].

\bibitem{Assel:2018vtq}
B.~Assel and A.~Tomasiello, {\it {Holographic duals of 3d S-fold CFTs}},  {\em
  JHEP} {\bf 06} (2018) 019, [\href{http://arxiv.org/abs/1804.06419}{{\tt
  arXiv:1804.06419}}].

\bibitem{Guarino:2020gfe}
A.~Guarino, C.~Sterckx, and M.~Trigiante, {\it {$\mathcal{N}=2$ supersymmetric
  S-folds}},  {\em JHEP} {\bf 04} (2020) 050,
  [\href{http://arxiv.org/abs/2002.03692}{{\tt arXiv:2002.03692}}].

\bibitem{Giambrone:2021zvp}
A.~Giambrone, E.~Malek, H.~Samtleben, and M.~Trigiante, {\it {Global properties
  of the conformal manifold for S-fold backgrounds}},  {\em JHEP} {\bf 06}
  (2021), no.~111 111, [\href{http://arxiv.org/abs/2103.10797}{{\tt
  arXiv:2103.10797}}].

\bibitem{Guarino:2021kyp}
A.~Guarino and C.~Sterckx, {\it {S-folds and holographic RG flows on the
  D3-brane}},  {\em JHEP} {\bf 06} (2021) 051,
  [\href{http://arxiv.org/abs/2103.12652}{{\tt arXiv:2103.12652}}].

\bibitem{Arav:2021gra}
I.~Arav, J.~P. Gauntlett, M.~M. Roberts, and C.~Rosen, {\it {Marginal
  deformations and RG flows for type IIB S-folds}},  {\em JHEP} {\bf 07} (2021)
  151, [\href{http://arxiv.org/abs/2103.15201}{{\tt arXiv:2103.15201}}].

\bibitem{Bobev:2021yya}
N.~Bobev, F.~F. Gautason, and J.~van Muiden, {\it {The holographic conformal
  manifold of 3d $ \mathcal{N} $ = 2 S-fold SCFTs}},  {\em JHEP} {\bf 07}
  (2021), no.~221 221, [\href{http://arxiv.org/abs/2104.00977}{{\tt
  arXiv:2104.00977}}].

\bibitem{Cesaro:2021tna}
M.~Ces\`aro, G.~Larios, and O.~Varela, {\it {The spectrum of
  marginally-deformed $ \mathcal{N} $ = 2 CFTs with AdS$_{4}$ S-fold duals of
  type IIB}},  {\em JHEP} {\bf 12} (2021) 214,
  [\href{http://arxiv.org/abs/2109.11608}{{\tt arXiv:2109.11608}}].

\bibitem{Guarino:2021hrc}
A.~Guarino and C.~Sterckx, {\it {Flat deformations of type IIB S-folds}},  {\em
  JHEP} {\bf 11} (2021) 171, [\href{http://arxiv.org/abs/2109.06032}{{\tt
  arXiv:2109.06032}}].

\bibitem{Bobev:2021rtg}
N.~Bobev, F.~F. Gautason, and J.~van Muiden, {\it {Holographic 3d $ \mathcal{N}
  $ = 1 conformal manifolds}},  {\em JHEP} {\bf 07} (2023) 220,
  [\href{http://arxiv.org/abs/2111.11461}{{\tt arXiv:2111.11461}}].

\bibitem{Guarino:2022tlw}
A.~Guarino and C.~Sterckx, {\it {Type IIB S-folds: flat deformations,
  holography and stability}},  {\em PoS} {\bf CORFU2021} (2022) 163,
  [\href{http://arxiv.org/abs/2204.09993}{{\tt arXiv:2204.09993}}].

\bibitem{Hull:1994ys}
C.~M. Hull and P.~K. Townsend, {\it {Unity of superstring dualities}},  {\em
  Nucl. Phys. B} {\bf 438} (1995) 109--137,
  [\href{http://arxiv.org/abs/hep-th/9410167}{{\tt hep-th/9410167}}].

\bibitem{Andrianopoli:2010bj}
L.~Andrianopoli, R.~D'Auria, S.~Ferrara, and M.~Trigiante, {\it {Fake
  Superpotential for Large and Small Extremal Black Holes}},  {\em JHEP} {\bf
  08} (2010) 126, [\href{http://arxiv.org/abs/1002.4340}{{\tt
  arXiv:1002.4340}}].

\bibitem{Helgason}
S.~Helgason, {\it {Differential Geometry, Lie Groups, and Symmetric Spaces}},
  {\em American Mathematical Society (First Edition, 2001).}

\bibitem{Bak:2003jk}
D.~Bak, M.~Gutperle, and S.~Hirano, {\it {A Dilatonic deformation of AdS(5) and
  its field theory dual}},  {\em JHEP} {\bf 05} (2003) 072,
  [\href{http://arxiv.org/abs/hep-th/0304129}{{\tt hep-th/0304129}}].

\bibitem{Clark:2004sb}
A.~B. Clark, D.~Z. Freedman, A.~Karch, and M.~Schnabl, {\it {Dual of the Janus
  solution: An interface conformal field theory}},  {\em Phys. Rev. D} {\bf 71}
  (2005) 066003, [\href{http://arxiv.org/abs/hep-th/0407073}{{\tt
  hep-th/0407073}}].

\bibitem{DHoker:2007zhm}
E.~D'Hoker, J.~Estes, and M.~Gutperle, {\it {Exact half-BPS Type IIB interface
  solutions. I. Local solution and supersymmetric Janus}},  {\em JHEP} {\bf 06}
  (2007) 021, [\href{http://arxiv.org/abs/0705.0022}{{\tt arXiv:0705.0022}}].

\bibitem{DHoker:2007hhe}
E.~D'Hoker, J.~Estes, and M.~Gutperle, {\it {Exact half-BPS Type IIB interface
  solutions. II. Flux solutions and multi-Janus}},  {\em JHEP} {\bf 06} (2007)
  022, [\href{http://arxiv.org/abs/0705.0024}{{\tt arXiv:0705.0024}}].

\bibitem{Dabholkar:2002sy}
A.~Dabholkar and C.~Hull, {\it {Duality twists, orbifolds, and fluxes}},  {\em
  JHEP} {\bf 09} (2003) 054, [\href{http://arxiv.org/abs/hep-th/0210209}{{\tt
  hep-th/0210209}}].

\bibitem{Schwarz:1983qr}
J.~H. Schwarz, {\it {Covariant Field Equations of Chiral N=2 D=10
  Supergravity}},  {\em Nucl. Phys. B} {\bf 226} (1983) 269.

\bibitem{Dorey:2001qj}
N.~Dorey, T.~J. Hollowood, and S.~P. Kumar, {\it {An Exact elliptic
  superpotential for N=1* deformations of finite N=2 gauge theories}},  {\em
  Nucl. Phys. B} {\bf 624} (2002) 95--145,
  [\href{http://arxiv.org/abs/hep-th/0108221}{{\tt hep-th/0108221}}].

\bibitem{Tanii:1984zk}
Y.~Tanii, {\it {$N=8$ Supergravity in Six-dimensions}},  {\em Phys. Lett. B}
  {\bf 145} (1984) 197--200.

\bibitem{Seiberg:1999xz}
N.~Seiberg and E.~Witten, {\it {The D1 / D5 system and singular CFT}},  {\em
  JHEP} {\bf 04} (1999) 017, [\href{http://arxiv.org/abs/hep-th/9903224}{{\tt
  hep-th/9903224}}].

\bibitem{Andrianopoli:1996zg}
L.~Andrianopoli, R.~D'Auria, S.~Ferrara, P.~Fre, R.~Minasian, and M.~Trigiante,
  {\it {Solvable Lie algebras in type IIA, type IIB and M theories}},  {\em
  Nucl. Phys. B} {\bf 493} (1997) 249--280,
  [\href{http://arxiv.org/abs/hep-th/9612202}{{\tt hep-th/9612202}}].

\bibitem{Cremmer:1997ct}
E.~Cremmer, B.~Julia, H.~Lu, and C.~N. Pope, {\it {Dualization of dualities.
  1.}},  {\em Nucl. Phys. B} {\bf 523} (1998) 73--144,
  [\href{http://arxiv.org/abs/hep-th/9710119}{{\tt hep-th/9710119}}].

\bibitem{Hull:2020byc}
C.~Hull, E.~Marcus, K.~Stemerdink, and S.~Vandoren, {\it {Black holes in string
  theory with duality twists}},  {\em JHEP} {\bf 07} (2020) 086,
  [\href{http://arxiv.org/abs/2003.11034}{{\tt arXiv:2003.11034}}].

\bibitem{Strominger:1996sh}
A.~Strominger and C.~Vafa, {\it {Microscopic origin of the Bekenstein-Hawking
  entropy}},  {\em Phys. Lett. B} {\bf 379} (1996) 99--104,
  [\href{http://arxiv.org/abs/hep-th/9601029}{{\tt hep-th/9601029}}].

\bibitem{Cremmer:1979uq}
E.~Cremmer, J.~Scherk, and J.~H. Schwarz, {\it {Spontaneously Broken N=8
  Supergravity}},  {\em Phys. Lett. B} {\bf 84} (1979) 83--86.

\bibitem{deWit:2002vt}
B.~de~Wit, H.~Samtleben, and M.~Trigiante, {\it {On Lagrangians and gaugings of
  maximal supergravities}},  {\em Nucl. Phys. B} {\bf 655} (2003) 93--126,
  [\href{http://arxiv.org/abs/hep-th/0212239}{{\tt hep-th/0212239}}].

\bibitem{Andrianopoli:2004xu}
L.~Andrianopoli, S.~Ferrara, and M.~A. Lledo, {\it {No-scale D=5 supergravity
  from Scherk-Schwarz reduction of D=6 theories}},  {\em JHEP} {\bf 06} (2004)
  018, [\href{http://arxiv.org/abs/hep-th/0406018}{{\tt hep-th/0406018}}].

\bibitem{Gkountoumis:2023fym}
G.~Gkountoumis, C.~Hull, K.~Stemerdink, and S.~Vandoren, {\it {Freely acting
  orbifolds of type IIB string theory on T$^{5}$}},  {\em JHEP} {\bf 08} (2023)
  089, [\href{http://arxiv.org/abs/2302.09112}{{\tt arXiv:2302.09112}}].

\bibitem{Bobev:2020fon}
N.~Bobev, F.~F. Gautason, K.~Pilch, M.~Suh, and J.~van Muiden, {\it
  {Holographic interfaces in $ \mathcal{N} $ = 4 SYM: Janus and J-folds}},
  {\em JHEP} {\bf 05} (2020) 134, [\href{http://arxiv.org/abs/2003.09154}{{\tt
  arXiv:2003.09154}}].

\bibitem{Chiodaroli:2009yw}
M.~Chiodaroli, M.~Gutperle, and D.~Krym, {\it {Half-BPS Solutions locally
  asymptotic to AdS(3) x S**3 and interface conformal field theories}},  {\em
  JHEP} {\bf 02} (2010) 066, [\href{http://arxiv.org/abs/0910.0466}{{\tt
  arXiv:0910.0466}}].

\bibitem{Astesiano:2023iql}
D.~Astesiano and F.~F. Gautason, {\it {Supersymmetric wormholes in String
  theory}},  \href{http://arxiv.org/abs/2309.02481}{{\tt arXiv:2309.02481}}.

\bibitem{Ferrara:1996um}
S.~Ferrara and R.~Kallosh, {\it {Universality of supersymmetric attractors}},
  {\em Phys. Rev. D} {\bf 54} (1996) 1525--1534,
  [\href{http://arxiv.org/abs/hep-th/9603090}{{\tt hep-th/9603090}}].

\bibitem{Ferrara:1997tw}
S.~Ferrara, G.~W. Gibbons, and R.~Kallosh, {\it {Black holes and critical
  points in moduli space}},  {\em Nucl. Phys. B} {\bf 500} (1997) 75--93,
  [\href{http://arxiv.org/abs/hep-th/9702103}{{\tt hep-th/9702103}}].

\bibitem{Larsen:1999uk}
F.~Larsen and E.~J. Martinec, {\it {U(1) charges and moduli in the D1 - D5
  system}},  {\em JHEP} {\bf 06} (1999) 019,
  [\href{http://arxiv.org/abs/hep-th/9905064}{{\tt hep-th/9905064}}].

\end{thebibliography}\endgroup
\bibliographystyle{JHEP}

\end{document}